# A globally convergent algorithm for lasso-penalized mixture of linear regression models

Luke R. Lloyd-Jones[1, *], Hien D. Nguyen[2, 3], and Geoffrey J. McLachlan[2]

[1]Centre for Neurogenetics and Statistical Genomics, Queensland Brain Institute, University of Queensland, St Lucia, Qld, 4072, Australia
[2]School of Mathematics and Physics, University of Queensland, St Lucia, Qld, 4072, Australia
[3]Centre for Advanced Imaging, University of Queensland, St Lucia, Qld, 4072, Australia
[*]Corresponding author. E-mail: l.lloydjones@uq.edu.au

May 3, 2016

**Abstract**

Variable selection is an old and pervasive problem in regression analysis. One solution is to impose a lasso penalty to shrink parameter estimates toward zero and perform continuous model selection. The lasso-penalized mixture of linear regressions model (L-MLR) is a class of regularization methods for the model selection problem in the fixed number of variables setting. In this article, we propose a new algorithm for the maximum penalized-likelihood estimation of the L-MLR model. This algorithm is constructed via the minorization–maximization algorithm paradigm. Such a construction allows for coordinate-wise updates of the parameter components, and produces globally convergent sequences of estimates that generate monotonic sequences of penalized log-likelihood values. These three features are missing in the previously presented approximate expectation-maximization algorithms. The previous difficulty in producing a globally convergent algorithm for the maximum penalized-likelihood estimation of the L-MLR model is due to the intractability of finding exact updates for the mixture model mixing proportions in the maximization-step. In our algorithm, we solve this issue by showing that it can be converted into a polynomial root finding problem. Our solution to this problem involves a polynomial basis conversion that is interesting in its own right. The method is tested in simulation and with an application to Major League Baseball salary data from the 1990s and the present day. We explore the concept of whether player salaries are associated with batting performance.

# 1 Introduction

Variable selection is an old and pervasive problem in regression analysis and has been widely discussed because of this; see George (2000) and Greene (2003, Ch. 8) for classical introductions to the topic, and see Hastie et al. (2009, Ch. 3) and Izenman (2008, Ch. 5) for some modern perspectives. In recent years, regularization has become popular in the statistics and machine learning literature, stemming from the seminal paper of Tibshirani (1996) on the least absolute shrinkage and selection operator (lasso). A recent account of the literature regarding the lasso and related regularization methods can be found in Buhlmann and van de Geer (2011). The mixture of linear regressions (MLR) for modeling heterogeneous data was first considered in Quandt (1972). The introduction of the expectation–maximization (EM) algorithm by Dempster et al. (1977) made such models simpler to estimate in a practical setting. Subsequently, MLR models became more popular; see DeSarbo and Cron (1988), De Veaux (1989), and Jones and McLachlan (1992) for example.

The lasso-penalized MLR model (L-MLR) was considered in Khalili and Chen (2007) among a class of other regularization methods for the selection problem in the fixed number of variables setting. The L-MLR was then generalized to the divergent number of variables setting in Khalili and Lin (2013), and to the mixture of experts setting in Khalili (2010). Furthermore, Stadler et al. (2010) (see also Buhlmann and van de Geer (2011, Sec. 9.2)) considered an alternative parameterization of the L-MLR to Khalili and Chen (2007), and suggested a modified regularization expression. An alternative modified grouped lasso criterion (Yuan and Lin, 2006) was suggested for regularization of the MLR problem in Hui et al. (2015). A recent review of the literature regarding the variable selection problem in MLR models can be found in Khalili (2011).

In this article, we propose a new algorithm for the maximum penalized-likelihood (MPL) estimation of L-MLR models. This algorithm is constructed via the MM (minorization–maximization) algorithm paradigm of Lange (2013, Ch. 8). Such a construction allows for some desirable features such as coordinate-wise updates of parameters, monotonicity of the penalized likelihood sequence, and global convergence of the estimates to a stationary point of the penalized log-likelihood function. These three features are missing in the approximate-EM algorithm presented in Khalili and Chen (2007). Previously, MM algorithms have been suggested for the regularization of regression models in Hunter and Li (2005), where they are noted to be numerically stable. Coordinate-wise updates of parameters in lasso-type problems was considered in Wu and Lange (2008), who also noted such updates to be fast and stable when compared to alternative algorithms. Furthermore, Stadler et al. (2010) also consider a coordinate-wise update scheme in their generalized EM algorithm, although the global convergence properties of the algorithm could only be established for the MPL estimation of a modified case of the L-MLR model with a simplified penalization function.

The difficulty in producing a globally convergent algorithm for the MPL estimation of the L-MLR model, which led both Khalili and Chen (2007) and Stadler et al. (2010) to utilize approximation schemes, is due to the intractability of the problem of updating the mixture model mixing proportions in the maximization-step of their respective algorithms. In our algorithm, we solve this issue by showing that it can be converted into a polynomial root finding problem. Our solution to this problem involves a polynomial basis conversion that is interesting in its own right. Aside from the new algorithm, we also consider the use of the L-MLR as a screening mechanism in a two-step procedure, as suggested in Buhlmann and van de Geer (2011, Sec. 2.5). Here, the L-MLR model is used to select the variable subset (step one) to include in a subsequent estimation of a MLR model (step two). This procedure allows for the adaptation of available asymptotic results for MLR models, such as those of Nguyen and McLachlan (2015), in order to obtain consistency and asymptotically normal parameter estimators. Optimization of the lasso tuning parameter vector $\boldsymbol{\lambda}$ via derivative free numerical methods is also explored as an alternative to exhaustive grid search.

To supplement the presented algorithm and procedures, we perform a set of simulation studies to demonstrate the capacity of our methodology. A user-friendly program that implements the proposed algorithm in C++ is also available at `https://github.com/lukelloydjones/` and is shown to be capable of handling reasonably large estimation problems. To compare the performance of the method on real data, we analyse the same data set on Major League Baseball (MLB) salaries presented in Khalili and Chen (2007). This allows for an initial comparison with the foundational work and an exploration of whether common measures of batting performance are good predictors of how much a batter is paid. This analysis is supplemented with a current data set from MLB seasons 2011-15, which allows for an investigation into how the distribution of salaries has changed and whether the same or new predictors are relevant. Baseball has always had a fascination with statistics, with baseball's link with statistics going back to the origins of the sport (Marchi and Albert, 2013). In particular, economists have long had great interest in the labor market and finances associated with MLB (Brown et al., 2015). Of notable fame is the Moneyball hypothesis (Lewis, 2004), which states that the ability of a player to get 'on base' was undervalued in the baseball labor market (before 2003) (Hakes and Sauer, 2006). Baseball statistics on player and team performance are some of the best of any sport especially in the modern era. Furthermore, baseball owners and players agree that playing performance is measurable and is associated with salary (Scully, 1974; Fullerton Jr and Peach, 2016). In this article we emphasize the application of our new methodology to these data in so far as there exists a statistical association, and believe that the implications of the performance salary association, with respect to MLB as a whole, have been explored in more depth elsewhere.

The article proceeds as follows. In Section 2, we introduce the L-MLR model and present the MM algorithm for its MPL estimation. In Section 3, we discuss the use of L-MLR models for statistical inference, and present the two-stage screening and estimation procedure. Section 4 outlines the algorithm's implementation. Simulation studies are then presented in Section 5, and we apply our method to data from salaries of batters from Major League Baseball (MLB) from the 1990s and present day in Section 6. Conclusions are then drawn in Section 7.

# 2 Mixture of Linear Regressions Model

Let $Y_1, ..., Y_n \in \mathbb{R}$ be an independent and identically distributed (IID) random sample that is dependent on corresponding covariate vectors $\boldsymbol{x}_1, ..., \boldsymbol{x}_n \in \mathbb{R}^p$, and let $Z_t$ be a latent categorical random variable $(t = 1, ..., n)$ such that $z_t \in \{1, ..., g\}$, where $\mathbb{P}(Z_t = i) = \pi_i > 0$ and $\sum_{i=1}^{g} \pi_i = 1$. The MLR model can be defined via the conditional probability density characterization

$$f(y_t \mid \boldsymbol{x}_t, Z_t = i;\ \boldsymbol{\theta}) = \phi\left(y_t;\ \alpha_i + \boldsymbol{x}_t^T \boldsymbol{\beta}_i, \sigma_i^2\right),$$

which implies the marginal probability density characterization

$$f\left(y_i \mid \boldsymbol{x}_t;\ \boldsymbol{\theta}\right)=\sum_{i=1}^{g}\pi_i\phi\left(y_t;\ \alpha_i+\boldsymbol{x}_t^T\boldsymbol{\beta}_i,\sigma_i^2\right). \tag{1}$$

Here $\phi\left(y;\ \mu,\sigma^2\right)$ is a normal density function with mean $\mu$ and variance $\sigma^2$, and we say that $\phi\left(y_t;\ \alpha_i+\boldsymbol{x}_t^T\boldsymbol{\beta}_i,\sigma_i^2\right)$ is the $i$th mixture component density. The vectors $\boldsymbol{\beta}_i=\left(\beta_{i1},...,\beta_{ip}\right)^T\in\mathbb{R}^p$, and scalars $\alpha_i\in\mathbb{R}$ and $\sigma_i^2>0$ are the specific regression coefficients, intercepts, and variances of the $i$th component density, respectively. We put all of the parameter components into the parameter vector $\boldsymbol{\theta}=\left(\boldsymbol{\pi}^T,\boldsymbol{\alpha}^T,\boldsymbol{\beta}^T,\boldsymbol{\sigma}^T\right)^T$, where $\boldsymbol{\pi}=\left(\pi_1,...,\pi_g\right)^T$, $\boldsymbol{\alpha}=\left(\alpha_1,...,\alpha_g\right)^T$, $\boldsymbol{\beta}=\left(\boldsymbol{\beta}_1^T,...,\boldsymbol{\beta}_g^T\right)^T$, and $\boldsymbol{\sigma}=\left(\sigma_1^2,...,\sigma_g^2\right)^T$.

We let $y_1,...,y_n$ with covariates $\boldsymbol{x}_1,...,\boldsymbol{x}_n$ be an observed random sample arising from an MLR with unknown parameter $\boldsymbol{\theta}_0=\left(\boldsymbol{\pi}_0^T,\boldsymbol{\alpha}_0^T,\boldsymbol{\beta}_0^T,\boldsymbol{\sigma}_0^T\right)^T$. If no additional assumptions are made regarding the nature of the regression coefficients $\boldsymbol{\beta}_0$ the parameter vector can be estimated by the ML estimator $\widetilde{\boldsymbol{\theta}}_n$, where $\widetilde{\boldsymbol{\theta}}_n$ is an appropriate local maximizer of the log-likelihood function for the MLR model

$$\mathcal{L}_n\left(\boldsymbol{\theta}\right)=\sum_{t=1}^{n}\log\sum_{i=1}^{g}\pi_i\phi\left(y_t;\ \alpha_i+\boldsymbol{x}_t^T\boldsymbol{\beta}_i,\sigma_i^2\right).$$

## 2.1 Lasso-penalized MLR

Suppose that it is known that $\boldsymbol{\beta}_0$ is sparse, in the sense that some or many elements of $\boldsymbol{\beta}_0$ are exactly equal to zero. The estimates for the zero elements of $\boldsymbol{\beta}_0$, obtained via $\widetilde{\boldsymbol{\theta}}_n$, will tend to be close to zero but will not be shrunked exactly to zero, and thus cannot be completely excluded from the model without the use of some other elimination techniques, such as hypothesis testing. One method for simultaneously shrinking insignificant regression coefficients to zero and estimating the parameter vector $\boldsymbol{\theta}_0$, as suggested by Khalili and Chen (2007), is to estimate the L-MLR by computing the MPL estimator $\widehat{\boldsymbol{\theta}}_n$, where $\widehat{\boldsymbol{\theta}}_n$ is an appropriate local maximizer of the lasso-penalized log-likelihood function for the MLR model

$$\mathcal{F}_n\left(\boldsymbol{\theta}\right)=\mathcal{L}_n\left(\boldsymbol{\theta}\right)-\mathcal{P}_n\left(\boldsymbol{\theta}\right). \tag{2}$$

Here

$$\mathcal{P}_n\left(\boldsymbol{\theta}\right)=\sum_{i=1}^{g}\pi_i\sum_{j=1}^{p}\lambda_{in}\mid\beta_{ij}\mid \tag{3}$$

is the mixture lasso penalty function, where $\lambda_{in} = n^{1/2}\gamma_{in}$ and $\gamma_{in} \geq 0$ are sequences of penalizing constants that are can be set to obtain a desired level of sparsity in the model. We note that $\widetilde{\boldsymbol{\theta}}_n$ and $\widehat{\boldsymbol{\theta}}_n$ are equivalent if $\lambda_{in} = 0$ for each $i$.

We now proceed to construct an MM algorithm for the MPL estimation of the L-MLR model. In order to produce an algorithm that is globally convergent, we follow the tactic of Hunter and Li (2005) and consider instead an appropriate local maximizer to the $\varepsilon$-approximate lasso-penalized log-likelihood function

$$\mathcal{F}_{n,\varepsilon}(\boldsymbol{\theta}) = \mathcal{L}_n(\boldsymbol{\theta}) - \mathcal{P}_{n,\varepsilon}(\boldsymbol{\theta}), \tag{4}$$

where

$$\mathcal{P}_{n,\varepsilon}(\boldsymbol{\theta}) = \sum_{i=1}^{g} \pi_i \sum_{j=1}^{p} \lambda_{in}\sqrt{\beta_{ij}^2 + \varepsilon^2} \tag{5}$$

for some small $\varepsilon > 0$. Similarly to Hunter and Li (2005, Prop. 3.2), we can show that $\mid \mathcal{F}_{n,\varepsilon}(\boldsymbol{\theta}) - \mathcal{F}_n(\boldsymbol{\theta}) \mid \to 0$ uniformly as $\varepsilon \to 0$, over any compact subset of the parameter space. The analysis of (4) instead of (2) is advantageous since its differentiability allows for the simple application of a useful global convergence theorem.

## 2.2 Minorization–Maximization Algorithm

We shall now proceed to describe the general framework of the block-wise MM algorithm. We note that these algorithms are also known as a block successive lower-bound maximization (BSLM) algorithm in the language of Razaviyayn et al. (2013).

Suppose that we wish to maximize some objective function $\mathcal{F}(\boldsymbol{\theta})$, where $\boldsymbol{\theta} = \left(\boldsymbol{\theta}_1^T, ..., \boldsymbol{\theta}_m^T\right)^T \in \Theta \subset \mathbb{R}^q$ and $\boldsymbol{\theta}_k \in \Theta_k \subset \mathbb{R}^{q_k}$, for $k = 1, ..., m$ and $\sum_{k=1}^{m} q_k = q$. If $\mathcal{F}(\boldsymbol{\theta})$ is difficult to maximize (e.g. the first order conditions of $\mathcal{F}$ are intractable, or $\mathcal{F}$ is non-differentiable), then we seek instead a sequence of simple iterates to maximize $\mathcal{F}$ instead.

We say that $\mathcal{U}_k(\boldsymbol{\theta}_k;\ \boldsymbol{\psi})$ is a minorizer of $\mathcal{F}$ over coordinate block $k$ if $\mathcal{F}(\boldsymbol{\psi}) = \mathcal{U}_k(\boldsymbol{\psi}_k;\ \boldsymbol{\psi})$ and $\mathcal{F}(\boldsymbol{\psi}) \geq \mathcal{U}(\boldsymbol{\theta}_k;\ \boldsymbol{\psi})$ whenever $\boldsymbol{\theta}_k \neq \boldsymbol{\psi}_k$. Here, $\boldsymbol{\psi} \in \Theta$ and $\boldsymbol{\psi}_k \in \Theta_k$. When $m = 1$, we say that $\mathcal{U}_1(\boldsymbol{\theta}_1;\ \boldsymbol{\psi}) = \mathcal{U}(\boldsymbol{\theta};\ \boldsymbol{\psi})$ minorizes $\mathcal{F}$. Upon finding an appropriate set of block-wise minorizers of $\mathcal{F}$, we can define a block-wise MM algorithm as follows.

Let $\boldsymbol{\theta}^{(0)}$ be some initialization and let $\boldsymbol{\theta}^{(r)}$ be the $r$th iterate of the algorithm. In the

$(r+1)$ th iteration, block-wise MM update scheme proceeds by computing

$$\boldsymbol{\theta}_k^{(r+1)} = \arg\max_{\boldsymbol{\theta}_k \in \Theta_k} \mathcal{U}\left(\boldsymbol{\theta}_k;\ \boldsymbol{\theta}^{(r)}\right)$$

for $k = (r \bmod m) + 1$, and setting $\boldsymbol{\theta}_l^{(r+1)} = \boldsymbol{\theta}_l^{(r)}$ for all $l \neq k$, where $l = 1, ..., m$. Given the definition of block-wise minorizers, it is not difficult to see that the iteration scheme that is described implies the monotonic ascent property

$$\mathcal{F}\left(\boldsymbol{\theta}^{(r+1)}\right) \geq \mathcal{U}_k\left(\boldsymbol{\theta}_k^{(r+1)};\ \boldsymbol{\theta}^{(r)}\right) \geq \mathcal{U}_k\left(\boldsymbol{\theta}_k^{(r)};\ \boldsymbol{\theta}^{(r)}\right) = \mathcal{F}\left(\boldsymbol{\theta}^{(r)}\right)$$

for every coordinate block $k$. We now present a set of useful minorizers for the MPL estimation of the L-MLR model.

**Lemma 1** *If $\theta, \varepsilon \in \mathbb{R}$, then the function $\mathcal{F}(\theta) = -\sqrt{\theta^2 + \varepsilon^2}$ can be minorized by*

$$\mathcal{U}(\theta;\ \psi) \quad = \quad -\frac{\theta^2}{2\sqrt{\psi^2 + \varepsilon^2}} + C(\psi),$$

*where $C(\psi) = -2^{-1}\sqrt{\psi^2 + \varepsilon^2} + \varepsilon^2\left(2\sqrt{\psi^2 + \varepsilon^2}\right)^{-1}$.*

**Lemma 2** *If $\Theta = [0, \infty)^q$, then the function $\mathcal{F}(\boldsymbol{\theta}) = \log\left(\sum_{i=1}^q \theta_i\right)$ can be minorized by the function*

$$\mathcal{U}(\boldsymbol{\theta};\ \boldsymbol{\psi}) = \sum_{i=1}^q \tau_i(\boldsymbol{\psi}) \log \theta_i + C(\boldsymbol{\psi}),$$

*where $\tau_i(\boldsymbol{\psi}) = \psi_i / \sum_{j=1}^q \psi_j$ and $C(\boldsymbol{\psi}) = -\sum_{i=1}^q \tau_i(\boldsymbol{\psi}) \log \tau_i(\boldsymbol{\psi})$.*

**Lemma 3** *If $\boldsymbol{\theta}, \boldsymbol{x} \in \mathbb{R}^q$ and $\alpha, y \in \mathbb{R}$, then the function $\mathcal{F}(\boldsymbol{\theta}) = -\left(y - \alpha - \boldsymbol{x}^T\boldsymbol{\theta}\right)^2$ can be minorized by*

$$\mathcal{U}(\boldsymbol{\theta};\ \boldsymbol{\psi}) = \frac{1}{q}\sum_{j=1}^q -\left[y - \alpha - q x_j\left(\theta_j - \psi_j\right) - \boldsymbol{x}^T\boldsymbol{\psi}\right]^2.$$

Lemmas 1 and 3 are adapted from the minorizers presented in Lange (2013, Sec. 8.5), and Lemma 2 can be found in Zhou and Lange (2010).

## 2.3 Derivation of Minorizers

Conditioned on the $r$th iterate $\boldsymbol{\theta}^{(r)}$, we can minorize $-\mathcal{P}_{n,\varepsilon}(\boldsymbol{\theta})$ by

$$\mathcal{U}_1\left(\boldsymbol{\theta};\ \boldsymbol{\theta}^{(r)}\right) = -\frac{1}{2}\sum_{i=1}^g \pi_i \sum_{j=1}^p \lambda_{in}\frac{\beta_{ij}^2}{w_{ij}^{(r)}} + C_1\left(\boldsymbol{\theta}^{(r)}\right), \tag{6}$$

where $w_{ij}^{(r)} = \sqrt{\beta_{ij}^{(r)2} + \varepsilon^2}$ and

$$C_1\left(\boldsymbol{\theta}^{(r)}\right) = -\frac{1}{2}\sum_{i=1}^{g}\pi_i\sum_{j=1}^{p}w_{ij}^{(r)} + \frac{\varepsilon^2}{2}\sum_{i=1}^{g}\pi_i\sum_{j=1}^{p}\frac{1}{w_{ij}^{(r)}},$$

by making the substitution $\theta = \beta_{ij}$ in Lemma 1, for each $i$ and $j$. We can write $\mathcal{U}_1\left(\boldsymbol{\theta};\ \boldsymbol{\theta}^{(r)}\right) = \mathcal{U}_1\left(\boldsymbol{\pi}, \boldsymbol{\beta};\ \boldsymbol{\theta}^{(r)}\right)$ since it does not depend on $\boldsymbol{\alpha}$ or $\boldsymbol{\sigma}$. Next, we can minorize $\mathcal{L}_n(\boldsymbol{\theta})$ by

$$\begin{aligned}\mathcal{U}_2\left(\boldsymbol{\theta};\ \boldsymbol{\theta}^{(r)}\right) &= \sum_{i=1}^{g}\sum_{t=1}^{n}\tau_{it}^{(r)}\left[\log\pi_i + \log\phi\left(y_t;\ \boldsymbol{x}_t^T\boldsymbol{\beta}_i, \sigma_i^2\right)\right] - \sum_{i=1}^{g}\sum_{t=1}^{n}\tau_{it}^{(r)}\log\tau_{it}^{(r)} \\ &= \sum_{i=1}^{g}\sum_{t=1}^{n}\tau_{it}^{(r)}\log\pi_i - \frac{1}{2}\sum_{i=1}^{g}\sum_{t=1}^{n}\tau_{it}^{(r)}\log\sigma_i^2 \\ &- \frac{1}{2}\sum_{i=1}^{g}\sum_{t=1}^{n}\frac{\tau_{it}^{(r)}}{\sigma_i^2}\left(y_t - \alpha_i - \boldsymbol{x}_t^T\boldsymbol{\beta}_i\right)^2 + C_2\left(\boldsymbol{\theta}^{(r)}\right)\end{aligned} \tag{7}$$

where $\tau_{it}^{(r)} = \pi_i\phi\left(y_t;\ \alpha_i + \boldsymbol{x}_t^T\boldsymbol{\beta}_i, \sigma_i^2\right)/f\left(y_i \mid \boldsymbol{x}_t;\ \boldsymbol{\theta}^{(r)}\right)$ and

$$C_2\left(\boldsymbol{\theta}^{(r)}\right) = -\frac{n}{2}\log(2\pi) - \sum_{i=1}^{g}\sum_{t=1}^{n}\tau_{it}^{(r)}\log\tau_{it}^{(r)},$$

by making the substitution $\theta_i = \pi_i\phi\left(y_t;\ \alpha_i + \boldsymbol{x}_t^T\boldsymbol{\beta}_i, \sigma_i^2\right)$ in Lemma 2, for each $i$. We can further minorize $\mathcal{U}_2\left(\boldsymbol{\theta};\ \boldsymbol{\theta}^{(r)}\right)$ by

$$\begin{aligned}\widetilde{\mathcal{U}}_2\left(\boldsymbol{\theta};\ \boldsymbol{\theta}^{(r)}\right) &= \sum_{i=1}^{g}\sum_{t=1}^{n}\tau_{it}^{(r)}\log\pi_i - \frac{1}{2}\sum_{i=1}^{g}\sum_{t=1}^{n}\tau_{it}^{(r)}\log\sigma_i^2 \\ &- \frac{1}{2p}\sum_{i=1}^{g}\sum_{j=1}^{p}\sum_{t=1}^{n}\frac{\tau_{it}^{(r)}}{\sigma_i^2}\left(y_t - \alpha_i - px_{tj}\left(\beta_{ij} - \beta_{ij}^{(r)}\right) - \boldsymbol{x}_t^T\boldsymbol{\beta}_i^{(r)}\right)^2 \\ &+ C_2\left(\boldsymbol{\theta}^{(r)}\right),\end{aligned} \tag{8}$$

by making the substitutions $\alpha = \alpha_i$, $y = y_t$, $\boldsymbol{x} = \boldsymbol{x}_t$, and $\boldsymbol{\theta} = \boldsymbol{\beta}_i$ in Lemma 3, for each $i$, $j$, and $t$.

Using results (6)–(8), we can deduce the following block-wise minorizers for the $\varepsilon$-approximate lasso-penalized log-likelihood function.

**Proposition 1** *Conditioned on the $r$th iterate $\boldsymbol{\theta}^{(r)}$, (4) can be block-wise minorized in the coordinates of the parameter components $\boldsymbol{\pi}$, $\boldsymbol{\alpha}$ and $\boldsymbol{\sigma}$, and $\boldsymbol{\beta}$, via the minorizers*

$$\mathcal{U}_{\boldsymbol{\pi}}\left(\boldsymbol{\pi};\ \boldsymbol{\theta}^{(r)}\right) = \mathcal{U}_2\left(\boldsymbol{\pi}, \boldsymbol{\alpha}^{(r)}, \boldsymbol{\beta}^{(r)}, \boldsymbol{\sigma}^{(r)};\ \boldsymbol{\theta}^{(r)}\right) - \mathcal{P}_{n,\varepsilon}\left(\boldsymbol{\theta}^{(r)}\right), \tag{9}$$

$$\mathcal{U}_{\boldsymbol{\alpha},\boldsymbol{\sigma}}\left(\boldsymbol{\alpha},\boldsymbol{\sigma};\ \boldsymbol{\theta}^{(r)}\right)=\mathcal{U}_{2}\left(\boldsymbol{\pi}^{(r)},\boldsymbol{\alpha},\boldsymbol{\beta}^{(r)},\boldsymbol{\sigma};\ \boldsymbol{\theta}^{(r)}\right)-\mathcal{P}_{n,\varepsilon}\left(\boldsymbol{\theta}^{(r)}\right), \tag{10}$$

*and*

$$\mathcal{U}_{\boldsymbol{\beta}}\left(\boldsymbol{\beta};\ \boldsymbol{\theta}^{(r)}\right)\ =\ \mathcal{U}_{1}\left(\boldsymbol{\pi}^{(r)},\boldsymbol{\beta};\ \boldsymbol{\theta}^{(r)}\right)+\widetilde{\mathcal{U}}_{2}\left(\boldsymbol{\pi}^{(r)},\boldsymbol{\alpha}^{(r)},\boldsymbol{\beta},\boldsymbol{\sigma}^{(r)};\ \boldsymbol{\theta}^{(r)}\right), \tag{11}$$

*respectively.*

## 2.4 Maximization of Minorizers

We now seek to maximize (9) in the constraint set

$$\widetilde{\Theta}_{\boldsymbol{\pi}}=\left\{\boldsymbol{\pi}:\pi_{i}\geq\zeta,\sum_{i=1}^{g}\pi_{i}=1,i=1,...,g\right\},$$

for $\zeta<1/g$. This can be achieved by solving a Karush-Kuhn-Tucker (KKT) problem. For the current problem, the KKT theorem (cf. Nocedal and Wright (2006, Thm. 12.1)) states that if $\boldsymbol{\pi}^{*}$ is a local maximizer of (9) in $\widetilde{\Theta}_{\boldsymbol{\pi}}$ that satisfies some appropriate constraint qualifications, then there exists Lagrange multipliers $\xi^{*}\in\mathbb{R}$ and $\boldsymbol{\eta}^{*}=\left(\eta_{1}^{*},...,\eta_{g}^{*}\right)^{T}$, such that $\nabla_{\boldsymbol{\pi}}\Lambda\left(\boldsymbol{\pi}^{*},\xi^{*},\boldsymbol{\eta}^{*}\right)=\mathbf{0}$ (KKT1), $\sum_{i=1}^{g}\pi_{i}=1$ (KKT2), $\pi_{i}^{*}\geq0$ for each $i$ (KKT3), $\eta_{i}^{*}\geq0$ for each $i$ (KKT4), and $\eta_{i}^{*}\left(\zeta-\pi_{i}^{*}\right)=0$ for each $i$ (KKT5). Here,

$$\Lambda\left(\boldsymbol{\pi},\xi,\boldsymbol{\eta}\right)=\mathcal{U}_{\boldsymbol{\pi}}\left(\boldsymbol{\pi};\ \boldsymbol{\theta}^{(r)}\right)+\sum_{i=1}^{g}\eta_{i}\left(\zeta-\pi_{i}\right)+\xi\left(\sum_{i=1}^{g}\pi_{i}-1\right)$$

is the KKT Lagrangian, $\nabla_{\boldsymbol{\pi}}\Lambda$ is the partial derivative of $\Lambda$ in $\boldsymbol{\pi}$, and $\mathbf{0}$ is a zero vector of appropriate dimensionality.

Firstly, we note that (9) is concave with respect to $\boldsymbol{\pi}$ and is defined over $\widetilde{\Theta}_{\boldsymbol{\pi}}$. Thus a global maximum in $\widetilde{\Theta}_{\boldsymbol{\pi}}$ must exist by the compactness of the set. Next, we observe that all of the constraints in $\widetilde{\Theta}_{\boldsymbol{\pi}}$ are linear and are thus qualified in the sense of the KKT theorem via Nocedal and Wright (2006, Lem. 12.7), thus a solution that satisfies KKT1–KKT5 must exist.

Suppose that a solution exist whereupon $\eta_{i}^{*}>0$ for some $i$. By KKT5, this implies that $\pi_{i}^{*}=\zeta$. Now, if we take the limit $\zeta\rightarrow0$, we observe that $\mathcal{U}_{\boldsymbol{\pi}}\left(\boldsymbol{\pi}^{*};\ \boldsymbol{\theta}^{(r)}\right)\rightarrow-\infty$, since $\log\zeta\rightarrow-\infty$, and so a solution cannot be the global maximum for a sufficiently small $\zeta$. We thus have the following result.

**Lemma 4** *There exists a global maximizer* $\boldsymbol{\pi}^*$ *of (9) in the set*

$$\Theta_{\boldsymbol{\pi}} = \left\{ \boldsymbol{\pi} : \pi_i > 0, \sum_{i=1}^{g} \pi_i = 1, i = 1, ..., g \right\}.$$

To obtain $\boldsymbol{\pi}^*$, we note that $\partial\Lambda/\partial\pi_i = \pi_i^{-1} a_i - b_i + \xi$, where $a_i = \sum_{t=1}^{n} \tau_{it}^{(r)}$ and $b_i = \lambda_{in} \sum_{j=1}^{p} \mid \beta_{ij}^{(r)} \mid$, for each $i$. We obtain the following result via KKT1 and KKT2.

**Theorem 1** *The global maximizer* $\boldsymbol{\pi}^*$ *of (9) in the set* $\Theta_{\boldsymbol{\pi}}$ *has the form*

$$\pi_i^* = \frac{a_i}{\xi^* - b_i} \tag{12}$$

*for each* $i = 1, ..., g$, *where* $\xi^* \in \Xi$ *and*

$$\Xi = \left\{ \xi : \sum_{i=1}^{g} \frac{a_i}{\xi - b_i} = 1 \right\}. \tag{13}$$

Via elementary algebra, we deduce that any equation (in $\xi$) of form (13) can be written as a polynomial of the form

$$\prod_{i=1}^{g} (\xi - b_i) - \sum_{i=1}^{g} a_i \prod_{j \neq i} (\xi - b_j) = 0. \tag{14}$$

For $g = 2$, (14) has the form

$$(\xi - b_1)(\xi - b_2) - a_1(\xi - b_2) - a_2(\xi - b_1) = 0,$$

which can be expanded and collected to yield

$$\xi^2 - (a_1 + a_2 + b_1 + b_2)\xi + (b_1 b_2 + a_1 b_2 + a_2 b_1) = 0.$$

The quadratic equation can then be used obtain the result that

$$\Xi = \left\{ \frac{1}{2}(a_1 + a_2 + b_1 + b_2) \pm \frac{1}{2}\sqrt{(a_1 + a_2 + b_1 + b_2)^2 - 4(b_1 b_2 + a_1 b_2 + a_2 b_1)} \right\}. \tag{15}$$

For $g = 3, 4$, it is more convenient to use a polynomial root finding algorithm rather than to solve (13) algebraically, and for $g \geq 5$ it is only possible to deduce $\Xi$ in this way; see McNamee (1993) and Pan (1997) for details regarding polynomial root finding. Unfortunately, (14) is not in the usual monomial basis form that is required by most algorithms. We resolve this issue via the transformation theorem of Gander (2005) to obtain the following result.

**Theorem 2** *Using the result in Appendix A, the left-hand side of (14) can be written in the monomial basis form*

$$Q_g\left(\xi\right)=\sum_{j=0}^{g-1}\left(c_j^{*}-\sum_{i=1}^{g}c_j^{(i)}\right)\xi^j+c_g^{*}\xi^g,$$

*where* $\mathbf{c}^{(i)}=\left(c_0^{(i)},...,c_{g-1}^{(i)}\right)^T$*, for* $i=1,...,g$*, and* $\boldsymbol{c}^{*}=\left(c_0^{*},...,c_g^{*}\right)^T$ *are as given in Appendix A.*

Using Theorems 1 and 2, and a suitable polynomial root finding algorithm, we can compute the $(r+1)$ th iterate block-wise update

$$\boldsymbol{\pi}^{(r+1)}=\arg\max_{\boldsymbol{\pi}\in\Theta_{\boldsymbol{\pi}}}\mathcal{U}_{\boldsymbol{\pi}}\left(\boldsymbol{\pi};\ \boldsymbol{\theta}^{(r)}\right) \tag{16}$$

by choosing the root $\xi^{*}\in\Xi$ that results in a $\boldsymbol{\pi}^{*}$ (via (12)) that lies in the set $\Theta_{\boldsymbol{\pi}}$, and that maximizes (9) among all other roots in $\Xi$.

Fortunately, the block-wise updates for $\boldsymbol{\alpha}$ and $\boldsymbol{\sigma}$ can be obtained from (10) via the first-order condition $\nabla\mathcal{U}_{\boldsymbol{\alpha}.\boldsymbol{\sigma}}\left(\boldsymbol{\alpha},\boldsymbol{\sigma};\ \boldsymbol{\theta}^{(r)}\right)=\mathbf{0}$. By doing so, we obtain the updates

$$\alpha_i^{(r+1)}=\frac{\sum_{t=1}^{n}\tau_{it}^{(r)}\left(y_t-\sum_{j=1}^{p}\beta_{ij}^{(r)}x_{tj}\right)}{\sum_{t=1}^{n}\tau_{it}^{(r)}} \tag{17}$$

and

$$\sigma_i^{2(r+1)}=\frac{\sum_{t=1}^{n}\tau_{it}^{(r)}\left(y_t-\alpha_i^{(r+1)}-\sum_{j=1}^{p}\beta_{ij}^{(r)}x_{tj}\right)^2}{\sum_{t=1}^{n}\tau_{it}^{(r)}} \tag{18}$$

for each $i$. Similarly, solving $\nabla\mathcal{U}_{\boldsymbol{\beta}}\left(\boldsymbol{\beta};\ \boldsymbol{\theta}^{(r)}\right)=\mathbf{0}$ yields the coordinate-wise updates for the $\boldsymbol{\beta}$ block

$$\beta_{ij}^{(r+1)}=\frac{p\sum_{t=1}^{n}\tau_{it}^{(r)}x_{tj}^2\beta_{ij}^{(r)}+\sum_{t=1}^{n}\tau_{it}^{(r)}x_{tj}\left(y_t-\alpha_i^{(r)}-\sum_{j=1}^{p}\beta_{ij}^{(r)}x_{tj}\right)}{\sigma_i^{2(r)}\pi_i^{(r)}\lambda_{in}w_{ij}^{(r)}+p\sum_{t=1}^{n}\tau_{it}^{(r)}x_{tj}^2}, \tag{19}$$

for each $i$ and $j$.

Note that (10) is concave in the alternative parameterization $\boldsymbol{\alpha}$ and $\widetilde{\boldsymbol{\sigma}}=\left(\log\sigma_1^2,...,\log\sigma_g^2\right)^T$, and thus (17) and (18) globally maximize (10) over the parameter space $\Theta_{\boldsymbol{\alpha}}\times\Theta_{\boldsymbol{\sigma}}=\mathbb{R}^g\times(0,\infty)^g$. Furthermore, (11) is concave in $\boldsymbol{\beta}$, therefore (19) globally maximizes (11) over $\Theta_{\boldsymbol{\beta}}=\mathbb{R}^{pg}$. The concavity of (10) and (11) can be deduced in a similar manner to the proof in Nguyen and McLachlan (2015, Thm. 2).

In summary, the block-wise MM algorithm for the maximization of (4) proceeds as follows. Initialize the algorithm with $\boldsymbol{\theta}^{(0)}$. Without loss of generality, at the $(r+1)$ th step of the algorithm, perform one of the following updates:

- If $(r \bmod 3)+1=1$, then perform the update (16) and set $\boldsymbol{\alpha}^{(r+1)}=\boldsymbol{\alpha}^{(r)}$, $\boldsymbol{\beta}^{(r+1)}=\boldsymbol{\beta}^{(r)}$, and $\boldsymbol{\sigma}^{(r+1)}=\boldsymbol{\sigma}^{(r)}$.

- If $(r \bmod 3)+1=2$, then perform updates (17) and (18) for each $i$, and set $\boldsymbol{\pi}^{(r+1)}=\boldsymbol{\pi}^{(r)}$ and $\boldsymbol{\beta}^{(r+1)}=\boldsymbol{\beta}^{(r)}$.

- If $(r \bmod 3)+1=3$, then perform the updates (19) for each $i$ and $j$, and set $\boldsymbol{\pi}^{(r+1)}=\boldsymbol{\pi}^{(r)}$, $\boldsymbol{\alpha}^{(r+1)}=\boldsymbol{\alpha}^{(r)}$, and $\boldsymbol{\sigma}^{(r+1)}=\boldsymbol{\sigma}^{(r)}$.

## 2.5 Convergence Analysis

The MM algorithm is iterated until some convergence criterion is met. In this article, we choose to use the absolute convergence criterion

$$\mathcal{F}_{n,\varepsilon}\left(\boldsymbol{\theta}^{(r+1)}\right)-\mathcal{F}_{n,\varepsilon}\left(\boldsymbol{\theta}^{(r)}\right)<\mathrm{TOL},$$

where $\mathrm{TOL}>0$ is a small tolerance constant; see Lange (2013, Sec. 11.4) for details regarding the relative merits of convergence criteria. Upon convergence, we denote the final iterate of the algorithm to be the MPL estimate and write it as $\widehat{\boldsymbol{\theta}}_n$.

Let $\boldsymbol{\theta}^*$ be a finite limit point of the MM algorithm, where $\boldsymbol{\theta}^*=\lim_{r\to\infty}\boldsymbol{\theta}^{(r)}$, or alternatively $\widehat{\boldsymbol{\theta}}_n\to\boldsymbol{\theta}^*$ as $\mathrm{TOL}\to 0$. Since each of the functions from Prop. 1 are concave over some bijective mapping of the parameters $\boldsymbol{\theta}$, and each of the updates (16)–(19) uniquely maximizer their respective block-wise minorizers, the MM algorithm is a BSLM algorithm that adheres to the assumptions of Razaviyayn et al. (2013, Thm. 2); we can inspect these properties in a similar manner to the proofs in Nguyen and McLachlan (2015, Thms. 2 and 3). We thus have the following result.

**Theorem 3** *If $\boldsymbol{\theta}^{(r)}$ is an iterate and $\boldsymbol{\theta}^*$ is a limit point of the block-wise MM algorithm for the MPL estimation of the L-MLR model (as given in Sec. 2.4) for some initialization $\boldsymbol{\theta}^{(0)}$, then the sequence $\mathcal{F}_{n,\varepsilon}\left(\boldsymbol{\theta}^{(r)}\right)$ is monotonically increasing in $r$, $\boldsymbol{\theta}^*$ is a coordinate-wise minimizer of (4), and $\boldsymbol{\theta}^*$ is a stationary point of (4).*

Theorem 3 is a useful result considering that (4) is both multimodal and unbounded. Because of this fact, the MM algorithm should be run multiple times from different initializations $\boldsymbol{\theta}^{(0)}$ in order to obtain an appropriate limit point. The issues of multimodality and local roots of mixture model likelihoods are discussed in McLachlan and Peel (2000, Sec. 2.12).

# 3 Statistical Inference

Let $\boldsymbol{x}_t$ be a realization of some random variable $\boldsymbol{X}_t$ with a well-behaved probability density $f(\boldsymbol{x}_t)$. Here, we leave the definition of well-behaved to be purposefully vague, as it is only required as a means of guaranteeing the uniform convergence of certain random functions. For argument sake, we can make $f(\boldsymbol{x}_t)$ well-behaved by supposing that it is a continuous density function defined over a compact domain.

Given an appropriate density function $f(\boldsymbol{x}_t)$, we can write the joint density of any pair $(\boldsymbol{X}_t, Y_t)$ as $f(\boldsymbol{x}_t, y_t;\ \boldsymbol{\theta}) = f(\boldsymbol{x}_t)\sum_{i=1}^{g}\pi_i\phi\left(y_t;\ \alpha_i + \boldsymbol{x}_t^T\boldsymbol{\beta}_i, \sigma_i^2\right)$ and interpret Khalili and Chen (2007, Thm. 1) as follows.

**Proposition 2** *If $(\boldsymbol{X}_t, Y_t)$, for $t = 1, ..., n$, is an IID random sample from a population with density function $f(\boldsymbol{x}_t, y_t;\ \boldsymbol{\theta}_0)$, where $f(\boldsymbol{x}_t)$ is well-behaved and $\boldsymbol{\theta}_0$ is the true value of the population parameter, then there exists a local maximizer $\widehat{\boldsymbol{\theta}}_n$ of (2) for which*

$$\left(\widehat{\boldsymbol{\theta}}_n - \boldsymbol{\theta}_0\right)^T\left(\widehat{\boldsymbol{\theta}}_n - \boldsymbol{\theta}_0\right) = O\left(n^{-1/2}\left[1 + \max_i \gamma_{in}\right]\right).$$

If we let $\gamma_{in} \to 0$ as $n \to \infty$ for each $i$, then Proposition 2 implies that the MPL estimators are root-$n$ consistent.

Unfortunately, as mentioned by (Khalili and Chen, 2007, Thm. 1), we cannot select sequences $\gamma_{in}$ such that $\widehat{\boldsymbol{\theta}}_n$ is both sparse and root-$n$ consistent simultaneously. That is, if we can decompose the population regression coefficients as $\boldsymbol{\beta}_{0i} = \left\{\boldsymbol{\beta}_{0i}^{[0]}, \boldsymbol{\beta}_{0i}^{[1]}\right\}$, for each $i$, where $\boldsymbol{\beta}_{0i}^{[1]}$ contains the non-zero coefficients and $\boldsymbol{\beta}_{0i}^{[0]} = \boldsymbol{0}$, then there is no choice of sequences $\gamma_{in}$ that would both make $\widehat{\boldsymbol{\theta}}_n$ root-$n$ consistent and make $\mathbb{P}\left(\widehat{\boldsymbol{\beta}}_n^{[0]} = \boldsymbol{0}\right) \to 1$ as $n \to \infty$. Here, we write $\boldsymbol{\beta}_0^{[k]} = \left(\boldsymbol{\beta}_{01}^{[k]T}, ..., \boldsymbol{\beta}_{0g}^{[k]T}\right)^T$ for $k = 0, 1$, and we write $\widehat{\boldsymbol{\beta}}_n^{[k]}$ as the MPL estimator of $\boldsymbol{\beta}_0^{[k]}$.

Furthermore, no limit point of the MM algorithm for maximizing (4) can be sparse, since the approximate penalty term (5) only allows for the shrinkage of coefficients in $\boldsymbol{\beta}^{[0]}$ down to a small constant proportional to $\varepsilon$. Because of these facts, we propose the following two-step procedure for obtaining sparse estimates of $\boldsymbol{\beta}_0$ using the MPL estimator of the L-MLR.

## 3.1 Variable Screening

As suggested in Buhlmann and van de Geer (2011, Sec. 2.5), we can select a small constant $C_\varepsilon$, depending on $\varepsilon$, that can be used to screen the variables obtained from the L-MLR estimator. That is, for each $i$ and $j$, we put coefficient $\beta_{ij}$ into $\boldsymbol{\beta}_i^{[0]}$ if $\mid \widehat{\beta}_{n,ij} \mid\leq C_\varepsilon$, and we put $\beta_{ij}$ into $\boldsymbol{\beta}_i^{[1]}$ otherwise. In this article, we choose to screen out any coefficient with absolute value in the order of magnitude of $\varepsilon$; that is we set $C_\varepsilon = \lfloor \log_{10} \varepsilon \rfloor$. This variable screening constitutes the first step of our two-step procedure.

In the second step of the procedure, we estimate the MLR model with all regression coefficients in $\boldsymbol{\beta}_i^{[0]}$ set to zero. That is, we estimate the model

$$f\left(y_i \mid \boldsymbol{x}_t;\ \boldsymbol{\theta}^{[1]}\right) = \sum_{i=1}^{g} \pi_i \phi\left(y_t;\ \alpha_i + \boldsymbol{x}_t^T \widetilde{\boldsymbol{\beta}}_i, \sigma_i^2\right), \tag{20}$$

where we can decompose the regression coefficients as $\widetilde{\boldsymbol{\beta}}_i = \left\{\boldsymbol{\beta}_i^{[0]}, \boldsymbol{\beta}_i^{[1]}\right\}$ with $\boldsymbol{\beta}_i^{[0]} = \mathbf{0}$. Here, $\boldsymbol{\theta}^{[1]} = \left(\boldsymbol{\pi}^T, \boldsymbol{\alpha}^T, \boldsymbol{\beta}^{[1]T}, \boldsymbol{\sigma}^T\right)^T$ is the parameter vector of (20), where $\boldsymbol{\beta}^{[1]} = \left(\boldsymbol{\beta}_1^{[1]T}, ..., \boldsymbol{\beta}_g^{[1]T}\right)^T$.

Upon defining the the ML estimator of $\boldsymbol{\theta}^{[1]}$ as an appropriate local maximizer of the likelihood function

$$\mathcal{L}_n^{[1]}\left(\boldsymbol{\theta}^{[1]}\right) = \sum_{t=1}^{n} \log \sum_{i=1}^{g} \pi_i \phi\left(y_t;\ \alpha_i + \boldsymbol{x}_t^T \widetilde{\boldsymbol{\beta}}_i, \sigma_i^2\right),$$

we can compute the ML estimate $\widetilde{\boldsymbol{\theta}}_n^{[1]}$ using the block-wise MM algorithm from Section 2.5 by setting $\lambda_{in} = 0$ for each $i$. We note that this implies that the update (16) reduces to the simplified form of $\pi_i^{(r+1)} = \sum_{t=1}^{n} \tau_{it}^{(r)}/n$ for each $i$.

Since $\widetilde{\boldsymbol{\theta}}_n^{[1]}$ is the ML estimate of an MLR, the usual asymptotic theorems for such models apply. The following results follow from minor modifications to Nguyen and McLachlan (2015, Thms. 5 and 6).

**Theorem 4** *Let $(\boldsymbol{X}_t, Y_t)$, for $t = 1, ..., n$, be an IID random sample from a population with density function $f(\boldsymbol{x}_t, y_t;\ \boldsymbol{\theta}_0) = f\left(\boldsymbol{x}_t, y_t;\ \boldsymbol{\theta}_0^{[1]}\right)$, where $f(\boldsymbol{x}_t)$ is well-behaved, $\boldsymbol{\theta}_0$ is the* *population parameter, and we can decompose the regression coefficients as* $\boldsymbol{\beta}_{0i} = \left\{\boldsymbol{\beta}_{0i}^{[0]}, \boldsymbol{\beta}_{0i}^{[1]}\right\}$ *with* $\boldsymbol{\beta}_{0i}^{[0]} = \mathbf{0}$*, and let* $\bar{\Theta}_n$ *be the set of roots of* $\nabla \mathcal{L}_n^{[1]}\left(\boldsymbol{\theta}^{[1]}\right) = \mathbf{0}$ *(*$\bar{\Theta}_n = \{\mathbf{0}\}$ *if no such root exists). If* $\boldsymbol{\theta}_0^{[1]}$ *is a local maximizer of* $\mathbb{E}\left[\log f\left(\boldsymbol{x}_t, y_t;\ \boldsymbol{\theta}^{[1]}\right)\right]$*, then for any* $\varepsilon > 0$*,*

$$\lim_{n\to\infty} \mathbb{P}\left[\inf_{\boldsymbol{\theta}^{[1]}\in\bar{\Theta}_n} \left(\widetilde{\boldsymbol{\theta}}_n^{[1]} - \boldsymbol{\theta}_0^{[1]}\right)^T \left(\widetilde{\boldsymbol{\theta}}_n^{[1]} - \boldsymbol{\theta}_0^{[1]}\right) > \varepsilon\right] = 0.$$

**Theorem 5** *Let* $\widetilde{\boldsymbol{\theta}}_n^{[1]}$ *be an ML estimator as in Theorem 4. If* $\boldsymbol{I}^{-1}\left(\boldsymbol{\theta}_0^{[1]}\right)$ *is positive definite, where*

$$\boldsymbol{I}\left(\boldsymbol{\theta}^{[1]}\right) = -\mathbb{E}\left[\frac{\partial^2 \log f\left(\boldsymbol{x}_t, y_t;\ \boldsymbol{\theta}^{[1]}\right)}{\partial\boldsymbol{\theta}^{[1]}\partial\boldsymbol{\theta}^{[1]T}}\right]$$

*is the Fisher information matrix, then* $n^{-1/2}\left(\widetilde{\boldsymbol{\theta}}_n^{[1]} - \boldsymbol{\theta}_0^{[1]}\right)$ *is asymptotically normal with mean vector* $\mathbf{0}$ *and covariance matrix* $\boldsymbol{I}^{-1}\left(\boldsymbol{\theta}_0^{[1]}\right)$*.*

Together, Theorems 4 and 5 can be used to obtain confidence intervals for our L-MLR screened regression coefficients. This is a useful result as it can be seen as an alternative to the usual oracle theorems for regularized regressions that depend on more complex penalization functions than the lasso; see for instance Khalili and Chen (2007, Thm. 2).

## 3.2 Tuning Parameter and Components Selection

In, Stadler et al. (2010) an information-theoretic method using the BIC (Bayesian information criterion) of Schwarz (1978) is suggested for the choice of tuning parameters $\boldsymbol{\lambda} = (\lambda_{1n}, ..., \lambda_{gn})^T$ and the number of components $g$ in the MLR model. We shall follow the same approach and describe our process as follows.

Suppose that $(g_0, \boldsymbol{\lambda}_0) \in \{(g_1, \boldsymbol{\lambda}_1), ..., (g_M, \boldsymbol{\lambda}_M)\}$ is a pair of parameters whereupon $g_0$ is the true number of components and $\boldsymbol{\lambda}_0$ results in the correct partitioning of regression coefficients, such that $\boldsymbol{\beta}_i^{[0]} = \boldsymbol{\beta}_{0i}^{[0]} = \mathbf{0}$. For each pair $(g_k, \boldsymbol{\lambda}_k)$, where $k = 1, ..., M$, we perform the two-step procedure described in Section 3.1 and obtain the parameter estimates $\widetilde{\boldsymbol{\theta}}_k^{[1]} = \left(\boldsymbol{\pi}_k^T, \boldsymbol{\alpha}_k^T, \boldsymbol{\beta}_k^{[1]T}, \boldsymbol{\sigma}_k^T\right)^T$ for a screened MLR model. The BIC for the model can then be computed as

$$\text{BIC}_k = -2\mathcal{L}_n^{[1]}\left(\widetilde{\boldsymbol{\theta}}_k^{[1]}\right) + (3g_k + p_k - 1)\log n, \tag{21}$$

where $p_k$ is the dimensionality of $\boldsymbol{\beta}_k^{[1]}$ (i.e. the total number of non-zero regression coefficients in the model). The BIC rule for tuning parameters selection is to set $(g, \boldsymbol{\lambda}) = \left(g_{\widetilde{k}}, \boldsymbol{\lambda}_{\widetilde{k}}\right)$, where

$$\widetilde{k} = \arg\min_{k} \text{BIC}_k.$$

We note that although the BIC has little theoretical support in the MLR setting, simulation studies such as those in Grun and Leisch (2007) and Nguyen and McLachlan (2016) indicate that the BIC tends to outperform or be comparable to other proposed criteria for such models.

# 4 Algorithm implementation

The proposed algorithm was implemented in the C++ programming language and a binary compiled for Mac OS X and source code are available at `https://github.com/lukelloydjones/`. The algorithm implementation is outlined in Section 4.1 (for when $g = 2$). The method requires the optimization over $\boldsymbol{\lambda} = (\lambda_1, \ldots, \lambda_g)^T$, which makes grid search prohibitively expensive for larger values of $g$. We use the ideas of Wu and Lange (2008), who suggest a shortcut to estimate $\lambda$ by combining bracketing and the golden section search algorithm (Kiefer, 1953). We advance this approach by using golden section search to initialize the $\boldsymbol{\lambda}$ vector and then optimize over all parameters in $\boldsymbol{\lambda}$ via the Nelder-Mead method (Nelder and Mead, 1965). This procedure begins with an initial definition of the bounds of the golden section search algorithm. All elements of the $\boldsymbol{\lambda}$ vector are then set to the current update of the golden section search algorithm and a minimum of equation (21) is determined for the initial one dimensional section $(\lambda_{b1}, \lambda_{b2})$. The elements of the $\boldsymbol{\lambda}$ vector are then set to the best $\lambda$ on $(\lambda_{b1}, \lambda_{b2})$, which initializes the starting simplex for the Nelder-Mead algorithm. The Nelder-Mead method then minimises (21) over $\boldsymbol{\lambda}$. We propose that this formulation is a heuristic that is practically more efficient than grid search for the multi-dimensional optimization of $\boldsymbol{\lambda}$.

## 4.1 Algorithm summary

### Algorithm 1

Initialize $\lambda_{b1}$, $\lambda_{b2}$, TOL, maximum iterations, $\varepsilon$, $\boldsymbol{\sigma}$, $\boldsymbol{\alpha}$, $\boldsymbol{\pi}$ from parameter file
Read in csv format $\boldsymbol{X}$ $(n \times p)$ and $\boldsymbol{y}$ $(n \times 1)$
**Golden section search algorithm start**. Note: sets $\lambda_i$ to be the a common $\lambda$ for all $i \in (1, \ldots, g)$
Set lower bound for golden search to $\lambda_{b1}$
Set upper bound for golden search to $\lambda_{b2}$
For initial $\lambda$ bounds run BIC optimization function and return minimum BIC using Eqn. (21)
**BIC optimization function start**
    Read in from file starting values for all $\boldsymbol{\beta}_i$
    Pass initial $\boldsymbol{\beta}_i$ to MLR optimization function
    **MLR optimization function start**
        Initialise $\boldsymbol{\lambda}, \boldsymbol{\beta_{0i}}$, TOL, maximum iterations, $\varepsilon$, $\boldsymbol{\sigma}$, $\boldsymbol{\alpha}$, $\boldsymbol{\pi}$ based on read parameters and current update of $\boldsymbol{\lambda}$
        *Initialize objective function*
        Set log-likelihood = 0.0
            for t :=1 to $n$ do
                for i :=1 to $g$ do
                $\mu_{ti} = \alpha_i + \boldsymbol{x}_t^T \boldsymbol{\beta_i}$
                likelihood += $\pi_i \phi(y_t;\, \mu_{ti}, \sigma_k)$ od
                log-likelihood += log(likelihood) od
            Subtract lasso penalisation
                for i :=1 to $g$ do
                log likelihood -= $\pi_k \lambda_k \sum_t \mid \beta_k \mid$ od
        *Initialize $\tau_i$ and $\omega_i$*
        *Begin main for loop*
        for iteration := 1 to maximum iterations do
            Update $\tau_i$
            for i :=1 to $g$ do
                $\tau_i = \pi_k \phi(y_t;\, \mu_{tk}, \sigma_k)/f(y_i \mid \boldsymbol{x}_t;\, \boldsymbol{\theta})$ do
            Calculate roots of KKT Lagrangian by Eqn. (15)
            Update $\pi_i$ to be the maximal set of roots in $(0, 1)$
            Update $\alpha_i$ by Eqn. (17)
            Update $\sigma_i$ by Eqn. (18)
            Update $\tau_i$ given new $\alpha_i$ and $\sigma_i$
            Update $\beta_{ij}$ by Eqn. (19)
            Update log likelihood as above
            if | log-likelihood − log-likelihood old | < TOL return optimal $\boldsymbol{\beta}_i$ od else continue fi
    **MLR optimization function end**
    If $\beta_{ij} < C_\varepsilon$, set $\beta_{ij} = 0$
    Run **MLR optimization function** with thresholded $\boldsymbol{\beta}_i$ and $\boldsymbol{\lambda} = \mathbf{0}$
    Return BIC from unpenalized **MLR optimization function**
**BIC optimization function end**
Iterate golden section search algorithm over $(\lambda_{b1}, \lambda_{b1})$ to find $\lambda$ that minimizes (21)
if upper bound − lower bound < 0.1 return optimal $\lambda$ and minimum BIC else continue fi
**Golden section search algorithm end**
**Downhill simplex method start**
    Initialize $\lambda_i$ simplex with best $\lambda$ from golden section search algorithm
    Downhill simplex method optimizes over BIC optimization function to find the best $\boldsymbol{\lambda}$ vector
**Downhill simplex method end**
Return best $\boldsymbol{\lambda}$ vector and BIC
Write $\boldsymbol{\beta}, \boldsymbol{\lambda}, \boldsymbol{\sigma}, \boldsymbol{\alpha}$, and $\boldsymbol{\pi}$ to file

# 5 Analysis of simulated data

To evaluate the performance of the proposed algorithm we simulated data under the normal MLR model with two components. We based our simulation model on that presented in Khalili and Chen (2007) and extended their set of $(n, p)$ scenarios using a similar structure to that presented in Wu and Lange (2008). Assuming that $g$ is known, the model for the first simulation was a $g = 2$ model of form (1) where $\pi_1 = 0.5$, $\alpha_1 = -20$, $\alpha_2 = 20$, $\sigma_1 = 1$, and $\sigma_2 = 1$. For each repetition of the simulation, columns of $\boldsymbol{X}$ were drawn from a binomial (2, $q$) distribution, with $q$ sampled from a uniform distribution on $(0.05, 0.5)$. This corresponds to columns with entries $0, 1$, or 2 with varying rates of success depending on the sampled $q$ for that column. This resembles data from genetics, where the values 0, 1, and 2 represent the counts of a reference allele for a gene or single nucleotide polymorphism (1000 Genomes Project Consortium and others, 2012). The methodology is to be used in this context in future work and thus we simulated $\boldsymbol{X}$ in this way. The matrix $\boldsymbol{X}$ is column standardised to have mean 0 and variance 1 before multiplication with $\boldsymbol{\beta}_i$. The vectors $\boldsymbol{\beta}_1$ and $\boldsymbol{\beta}_2$ had five non-zero elements each, with elements $1 \leqslant j \leqslant 5$ of $\boldsymbol{\beta}_1$ and the elements $5 \leqslant j \leqslant 10$ of $\boldsymbol{\beta}_2$ having an effect size of 5. Five scenarios were evaluated with varying values of $n$ and $p$ with 50 repititions done for each. Post analysis, the mean squared errors (MSE) between the predicted outcomes and the 'true' simulated outcomes from the L-MLR model were compared with those from multiple regression, or if $p > n$ marginal linear regression.

The second simulation used model (1) but with a much more challenging set of initial parameters, which included $\pi_1 = 0.3$, $\alpha_1 = 0$, $\alpha_2 = 7$, $\sigma_1 = 1$, and $\sigma_2 = 1$, and the matrix $\boldsymbol{X}$ simulated as in the first simulation. The number of non-zero elements of $\boldsymbol{\beta}_i$ varied, with elements $1 \leqslant j \leqslant 3$ of $\boldsymbol{\beta}_1$ equal to $(2, 1.5, 2)$ and elements $5 \leqslant j \leqslant 12$ of $\boldsymbol{\beta}_2$ equal to (0.5, -2, 1, 0.5, 0.5, 0.5, 2, -1). Again varying values of $n$ and $p$ were explored with 50 iterates performed for each of the $n$ and $p$ combinations. Post analysis the mean squared errors (MSE) from the predicted model were compared with those from multiple regression, or if $p > n$, marginal linear regression (i.e., simple-linear regression for each $j$).

Table 1 reports the results from the first simulation based on 50 replicates and combinations of the sample size $n$ and dimension $p$ specified by $(n, p) = (200, 20), (200, 100), (500, 200)$, $(200, 250)$, and $(500, 1000)$. For each of the scenarios $\boldsymbol{\beta}$ coefficients were estimated with high accuracy and precision. Reduced accuracy was seen when $p$ became larger than $n$ but precision remained approximately stable. The intercept terms $\boldsymbol{\alpha}$ and mixing proportions $\boldsymbol{\pi}$ were well estimated with high precision and accuracy seen for both. As $p$ increased the number of false non-zero elements ($N_{NZ}$) of $\boldsymbol{\beta}$ increased, with the mean for these parameters ranging from 0.04–0.19 across the scenarios, suggesting on average small effects for these predictors. In each scenario the MSE from the L-MLR model was substantially smaller than that from marginal or multiple regression. Average computing time in seconds (on an Intel Core i7 CPU running at 2.8 GHz with 16 GB internal RAM) was recorded in the last column of Table 1 showing a large increase in computing time for the $(n, p) = (500, 1000)$ case.

Table 2 reports results from the more challenging second simulation, which again contained 50 replicates across $(n, p) = (500, 20)$ and $(350, 400)$. The optimization routine incorporated two $\lambda$ coefficients, one for each of the regressor vectors that corresponded to the mixture components. For the $(n, p) = (500, 20)$ case, parameters were well estimated with low bias and high precision observed for the true non-zero elements of $\boldsymbol{\beta}$. For the $(n, p) = (350, 400)$ case, true non-zero elements of $\boldsymbol{\beta}$ were less well estimated with the true non-zero elements of $\boldsymbol{\beta}_1$ downwardly biased. For both simulation cases the intercept and mixing proportion parameters were well estimated with low bias and high precision seen for both. The number of false non-zero $\boldsymbol{\beta}$ elements increased with $p$ with mean values being 0.001 and $-0.007$ for each of the simulation cases respectively; again suggesting that although false non-zero elements are present the effect size is on average small. For the largest simulation scenario, computing time was in the order of half an hour with a standard deviation of approximately nine minutes.

Table 1: Summary of lasso-penalized mixture of linear regressions estimates and standard errors (below true values) from the first simulated scenario. Standard errors of estimates appear below parameter estimates. The number of false non-zero predictors ($N_{NZ}$) and the mean $\beta$ estimate over all $N_{NZ}$ are also reported. The mean squared error (MSE) for predictions from the L-MLR and multiple linear regression or, if $p > n$, marginal regression are reported for each of the simulations. Time is in seconds.

| $(n, p)$ | $\lambda$ | $\beta_{1,1}$ | $\beta_{1,2}$ | $\beta_{1,3}$ | $\beta_{1,4}$ | $\beta_{1,5}$ | $\beta_{2,6}$ | $\beta_{2,7}$ | $\beta_{2,8}$ | $\beta_{2,9}$ | $\beta_{2,10}$ | $\alpha_1$ | $\alpha_2$ | $\pi_1$ | $\pi_2$ | $N_{NZ}$ | Mean NZ | MSE$_{LMLR}$ | MSE$_R$ | Time |
|---|---|---|---|---|---|---|---|---|---|---|---|---|---|---|---|---|---|---|---|---|
| True values | | 5 | 5 | 5 | 5 | 5 | 5 | 5 | 5 | 5 | 5 | -20 | 20 | 0.5 | 0.5 | | | | | |
| (200, 20) | 6.2 | 5.00 | 5.02 | 5.03 | 4.99 | 5.02 | 4.98 | 5.02 | 5.00 | 5.00 | 5.04 | -20.0 | 20.0 | 0.499 | 0.501 | 10.5 | 0.10 | 1.2 | 414.9 | 1.4 |
| | 2.1 | 0.12 | 0.10 | 0.11 | 0.10 | 0.09 | 0.10 | 0.11 | 0.12 | 0.13 | 0.11 | 0.10 | 0.10 | 0.002 | 0.002 | 2.6 | 0.02 | 0.14 | 15.8 | 0.3 |
| (200, 100) | 36.1 | 5.00 | 4.97 | 4.97 | 4.99 | 4.97 | 4.97 | 4.99 | 4.99 | 4.97 | 4.98 | -20.0 | 20.0 | 0.500 | 0.500 | 4.5 | 0.19 | 1.1 | 230.1 | 12.7 |
| | 2.1 | 0.12 | 0.11 | 0.11 | 0.08 | 0.11 | 0.12 | 0.13 | 0.12 | 0.11 | 0.09 | 0.09 | 0.09 | 0.003 | 0.003 | 3.24 | 0.06 | 0.14 | 24.7 | 2.9 |
| (500, 200) | 48.6 | 4.98 | 5.00 | 4.98 | 4.98 | 4.97 | 4.99 | 4.98 | 4.99 | 4.98 | 4.98 | -20.0 | 20.0 | 0.500 | 0.500 | 29.4 | 0.11 | 1.2 | 279.4 | 106 |
| | 1.0 | 0.07 | 0.08 | 0.06 | 0.07 | 0.06 | 0.06 | 0.07 | 0.08 | 0.06 | 0.08 | 0.05 | 0.06 | 0.002 | 0.002 | 8.2 | 0.01 | 0.10 | 16.9 | 20 |
| (200, 250) | 40.9 | 4.97 | 4.98 | 4.96 | 4.99 | 4.98 | 4.95 | 4.97 | 4.98 | 4.97 | 4.97 | -19.9 | 20.0 | 0.500 | 0.500 | 19.7 | 0.16 | 1.3 | 820.5 | 119.5 |
| | 3.2 | 0.11 | 0.12 | 0.11 | 0.11 | 0.12 | 0.15 | 0.13 | 0.13 | 0.13 | 0.13 | 0.12 | 0.11 | 0.002 | 0.002 | 48.7 | 0.12 | 0.61 | 130.5 | 99.5 |
| (500, 1000) | 37.1 | 4.85 | 5.05 | 4.80 | 5.04 | 4.81 | 5.05 | 4.83 | 5.07 | 4.83 | 5.05 | -20.0 | 20.0 | 0.500 | 0.500 | 564.7 | 0.04 | 28.0 | 1793.4 | 39,918 |
| | 2.1 | 0.11 | 0.10 | 0.08 | 0.11 | 0.09 | 0.13 | 0.11 | 0.10 | 0.10 | 0.11 | 0.09 | 0.09 | 0.002 | 0.002 | 17.1 | 0.002 | 10.6 | 196.3 | 27,454 |

Table 2: Summary of lasso-penalized mixture of linear regressions estimates and standard errors from the second simulated scenario. Standard errors of estimates appear below parameter estimates. The number of falsely identified non-zero predictors ($N_{NZ}$) and the mean $\beta$ estimate over all $N_{NZ}$ are reported. The mean squared error (MSE) for predictions from the L-MLR and multiple linear regression or, if $p > n$, marginal regression are reported for each of the simulations. Time is in seconds.

| $(n, p)$ | $\lambda_1$ | $\lambda_2$ | $\beta_{1,1}$ | $\beta_{1,2}$ | $\beta_{1,3}$ | $\beta_{2,5}$ | $\beta_{2,6}$ | $\beta_{2,7}$ | $\beta_{2,8}$ | $\beta_{2,9}$ | $\beta_{2,10}$ | $\beta_{2,11}$ | $\beta_{2,12}$ | $\alpha_1$ | $\alpha_2$ | $\pi_1$ | $\pi_2$ | $N_{NZ}$ | Mean NZ | $\text{MSE}_{LMLR}$ | $\text{MSE}_R$ | Time |
|---|---|---|---|---|---|---|---|---|---|---|---|---|---|---|---|---|---|---|---|---|---|---|
| True values | | | 2 | 1.5 | 2 | 0.5 | -2 | 1 | 0.5 | 0.5 | 0.5 | 2 | -1 | 0 | 7 | 0.3 | 0.7 | | | | | |
| (500, 20) | 14.0 | 15.3 | 1.99 | 1.49 | 2.01 | 0.50 | -2.00 | 1.00 | 0.50 | 0.50 | 0.49 | 2.00 | -1.00 | 0.01 | 7.0 | 0.30 | 0.70 | 17.4 | 0.001 | 0.98 | 14.8 | 16.3 |
| | 3.4 | 4.5 | 0.11 | 0.10 | 0.08 | 0.06 | 0.06 | 0.06 | 0.05 | 0.05 | 0.06 | 0.06 | 0.05 | 0.08 | 0.06 | 0.01 | 0.01 | 3.4 | 0.02 | 0.15 | 2.2 | 3.1 |
| (350, 400) | 59.8 | 59.0 | 1.87 | 1.40 | 1.88 | 0.47 | -1.98 | 0.99 | 0.47 | 0.47 | 0.47 | 1.96 | -0.96 | -0.04 | 7.0 | 0.31 | 0.69 | 98.0 | -0.007 | 1.51 | 56.4 | 2193.1 |
| | 5.6 | 5.2 | 0.10 | 0.12 | 0.11 | 0.10 | 0.08 | 0.09 | 0.10 | 0.10 | 0.09 | 0.07 | 0.07 | 0.14 | 0.06 | 0.01 | 0.01 | 42.3 | 0.04 | 0.284 | 5.1 | 524.5 |

# 6 Analysis of real data

As a further test of the methodology, we investigated baseball salaries from the Journal of Statistics Education (`www.amstat.org/publications/jse`) with the intention to infer whether measures of batting performance affect an individual's salary. This was done to provide a baseline comparison with the foundational work of Khalili and Chen (2007) and for comparison with more recent data. These data contained salaries for 337 MLB players (batters only) from the year 1992, and 16 measures of batting performance from the year 1991. Each of the MLB players participated in at least one game in both the 1991 and 1992 seasons. As highlighted by Khalili and Chen (2007), the log(salary) histogram shows multi-modality making it a good candidate for a response variable under the MLR model (Figure 1).

Each of the measures of batting performance are summarised in Table 3. The same set of interaction terms from Khalili and Chen (2007) was used in the analysis, which included a further 16 interaction terms, making in total 32 predictors. Columns of $\boldsymbol{X}$ were standardised to have mean 0 and variance 1 for use with the methodology. For comparison, the `stepAIC` function from the `MASS` package (Venables and Ripley, 2002) of the R programming language (R Core Team, 2015) was used to implement model selection via the BIC for the standard linear model. For ease of comparison we reported the parameter estimates for $\boldsymbol{\beta}_1$ and $\boldsymbol{\beta}_2$ from the MIXLASSO method presented in Khalili and Chen (2007) with predictions from these values calculated assuming that $\boldsymbol{X}$ is column standardised (mean 0 and variance 1). Prediction with an unstandardised matrix yielded very poor results and thus we assumed that column-wise normalisation was conducted, although it was not explicitly stated in Khalili and Chen (2007). The $\boldsymbol{\beta}$ parameter estimates from Khalili and Chen (2007), were also used as starting values for the implementation of our method.

The predicted logged salaries from the L-MLR model had a MSE of 0.08 and a regression ($Y_{\text{pred}}$ on $Y_{\text{obs}}$) $R^2$ of 0.94. The predicted logged salaries from the stepwise-BIC linear model showed a MSE of 0.27 and $R^2$ of 0.80. The predicted logged salaries from MIXLASSO had a MSE 0.58 and $R^2$ of 0.67. These results suggest that the L-MLR model had the smallest MSE and explained the largest proportion of variance for the baseball salary data from the 1991/92 MLB seasons.

A plot of the the fitted posterior probabilities of membership of the first component of the mixture $\tau_1$ on the observed logged salaries revealed that the L-MLR model partitioned the batters into two groups (Figure S2). These groups coincided with a group of batters that were low and highly paid, and a group of averaged paid batters. For the low and high paid set of batters, the predictors were more in line with a naive interpretation i.e., those batters that make many hits and are struck out less are more highly paid with hits and doubles being large positive predictors and strikeouts and stolen bases negative predictors (Table 3). In contrast, the average paid batters appeared to have salaries that were mostly well predicted from free agency eligibility and arbitration eligibility. But hits and runs batted in were also positive predictors for the average paid batters. The predicted values from the stepwise-BIC linear model showed multi-modality and was a better predictor than the MIXLASSO model. Free agent eligibility, arbitration eligibility, hits and runs batted in were also positive predictors for this model. Strikeouts and free agency were negative predictors. The predictors for this model were in line with those from L-MLR. However, the inference to which group these predictors belonged could not be made.

To update these analyses, we gathered data from modern MLB batters from the 2011-15 seasons. Salaries for individuals were acquired from the website `http://www.usatoday.com/sports/mlb/salaries/` for all players in the MLB over these seasons. Salary values are based on documents obtained from club officials from the MLB Players Association. Incentive clauses and deferred payments are not included in the value. Salary values do not include money paid or received in trades or for players who have been released. Measures of batting performance for these players were gathered from the website `http://espn.go.com/mlb/statistics`, which include standard, expanded, and 'sabermetric' batting statistics for all batters in the MLB over the 2011-15 seasons. These seasons were played under the new bargaining agreement established in 2011 between the 30 Major League Clubs and the Major League Baseball Players Association (Major League Baseball Players Association, 2011). For each season, statistics were matched to players salaries in the R programming language with a total of 37 predictors available for each batter. The logged salaries (in 1000s) were again used as a response variable for the analysis. The distribution of combined modern salaries had three modes, which include a spike around $600,000 suggesting a large number of MLB players on a base salary, and two broader density components (Figure S1). These

three modes are likely to reflect the salary structure across Major League Baseball, which is dependent on the 2011 agreement between the 30 Major League Clubs and the Major League Baseball Players Association. In Section VI of this agreement the rules stipulate that the minimum salary for a player per season, in the MLB, shall be $480,000 to $520,000 for the 2012 to 2016 seasons. Any Player with a total of three or more years of Major League service, but with less than six years of service, may submit the issue of the player's salary to final and binding arbitration without the consent of the club (Major League Baseball Players Association, 2011). Arbitration allows a player to enter into salary negotiations based on their performance, with the effect of eligibility for salary arbitration on player salaries often being dramatic (Abrams, 2010). At the salary arbitration stage the player's salary is set in comparison with other players around the league with similar performance statistics and major league (Abrams, 2010). Players with 6 or more years of Major League service who have not executed a contract for the next succeeding season are eligible to become a free agent. The progression to arbitration eligibility or free agent status is associated with a large increase in the capacity of a player to earn a much larger salary. This induces a correlation between time played in the major league and pay, which is well known in the economics literature surrounding baseball (Scully, 1974; Hakes and Sauer, 2006). It is not until players reach free agent status that they may negotiate contracts with any team in a truly competitive labor market, often leading to a further dramatic increase in salary (over arbitration status) (Brown et al., 2015). However, players of rare talent earn salaries apparently in excess of their relative contribution to the team, which introduces a potential non-linearity into the performance salary model (Scully, 1974).

When modeled, no predictors were seen for the spike for minimum wage players as there was little variation around this value. We therefore chose to eliminate this spike for each year analysed with a threshold set depending on the spike for that that year (approximately $665,000 to $1,800,000 (USD) depending on the season). This left data sets comprising $n = 110, 94, 100, 112,$ and 99 for the 2011-15 seasons respectively. Seasons were analysed separately using the L-MLR model and the stepwise-BIC linear model. As contracts for batters in the MLB often extend for many years (5-10 years) a combined analysis over the five year period was also performed. This was done by averaging statistics and salaries for players who were present in the MLB over all or some of the years. This resulted in a data set of 215 batters (excluding those less than 6.5 on the logged salary scale) from the 2011-15 seasons. The primary motivation for this last analysis was the long term contracts present in the MLB and to increase sample size.

Years 2011–2015 showed MSEs of (0.05, 0.07, 0.06, 0.14, 0.06) for L-MLR and (0.21, 0.19, 0.35, 0.50, 0.27) for the stepwise-BIC model, with similar patterns in $R^2$ seen across the seasons. Predicted densities showed a good characterisation of the multi-modality in the logged salaries for the L-MLR model, with the stepwise-BIC not being able to model this (Figure 2). Plotting the observed salaries versus $\tau_1$ for each of the years modeled, showed that the method partitions the salaries of modern batters into lower and higher salaries, which is in contrast to the salaries observed in the 1990s (Figure S2). Thus, the components of $\boldsymbol{\beta}$ contain predictors that explain variation for the lower paid batters and higher paid batters. Across seasons, predictors were not consistent for both the stepwise-BIC and L-MLR models.

The averaged analysis showed a MSE of 0.12 and $R^2$ of 0.84 for the L-MLR method, with the stepwise-BIC showing a MSE of 0.49 and $R^2$ of 0.32. Positive predictors for the highest paid batters (mean salary of $\approx$ 7.3 million dollars) included runs, runs batted in, stolen bases, total plate appearances, and intentional walks (Table 4). Negative predictors for the highest paid batters included doubles, strikeouts, wins against replacement, games played, sacrifice hits, and at bats per home run. For medium paid batters (mean salary of $\approx$ 1.9 million dollars) runs and triples were the most important positive predictors.

Table 3: Summary of stepwise-BIC (SW-BIC), lasso-penalized mixture of linear regressions, and MIXLASSO estimates from baseball data from the 1990s. Covariate acronyms are AVG - Batting Average, OBP - On Base Percentage, R - Runs, H - Hits, 2B - Doubles, 3B - Triples, HR - Home Runs, RBI - Runs Batted In, BB - Walks, SO - Strikeouts, SB - Stolen Bases, ERS - Errors, FAE - Free Agency Eligibility, FA - Free Agent in 1991/2, AE - Arbitration Eligibility, ARB - Arbitration in 1991/2.

| Covariates | SW-BIC | MIXLASSO Khalili and Chen (2007) | | L-MLR | |
|---|---|---|---|---|---|
| | | Component 1 | Component 2 | Component 1 | Component 2 |
| Intercept | 6.54 | 6.41 | 7.00 | 6.58 | 6.39 |
| AVG | – | – | -0.32 | – | 0.05 |
| OBP | – | – | 0.29 | – | – |
| R | – | – | -0.70 | – | – |
| H | 0.26 | 0.20 | 0.96 | 0.16 | 0.54 |
| 2B | – | – | – | – | 0.45 |
| 3B | – | – | – | – | – |
| HR | – | -0.19 | – | – | – |
| RBI | 0.30 | 0.26 | – | 0.13 | – |
| BB | – | – | – | 0.08 | – |
| SO | -0.14 | – | – | – | -0.34 |
| SB | – | – | – | 0.05 | -0.29 |
| ERS | – | – | – | – | – |
| FAE | 0.83 | 0.79 | 0.70 | 0.98 | – |
| FA | -0.18 | 0.72 | – | – | – |
| AE | 0.53 | 0.15 | 0.50 | 0.60 | – |
| ARB | – | – | -0.36 | – | -0.12 |
| AVG×FAE | – | -0.21 | – | – | – |
| AVG×FA | – | 0.63 | – | – | – |
| AVG×AE | – | 0.34 | – | – | – |
| AVG×ARB | -0.19 | – | – | – | – |
| R×FAE | – | – | – | – | – |
| R×FA | – | 0.14 | -0.38 | – | -0.07 |
| R×AE | – | – | – | 0.03 | – |
| R×ARB | – | -0.18 | 0.74 | – | – |
| HR×FAE | – | – | – | 0.05 | – |
| HR×FA | 0.11 | – | – | – | – |
| HR×AE | – | – | 0.34 | – | – |
| HR×ARB | – | – | – | 0.03 | – |
| RBI×FAE | – | 0.29 | -0.46 | 0.01 | – |
| RBI×FA | – | -0.14 | – | – | – |
| RBI×AE | – | – | – | 0.07 | – |
| RBI×ARB | 0.19 | – | – | – | – |

(a)

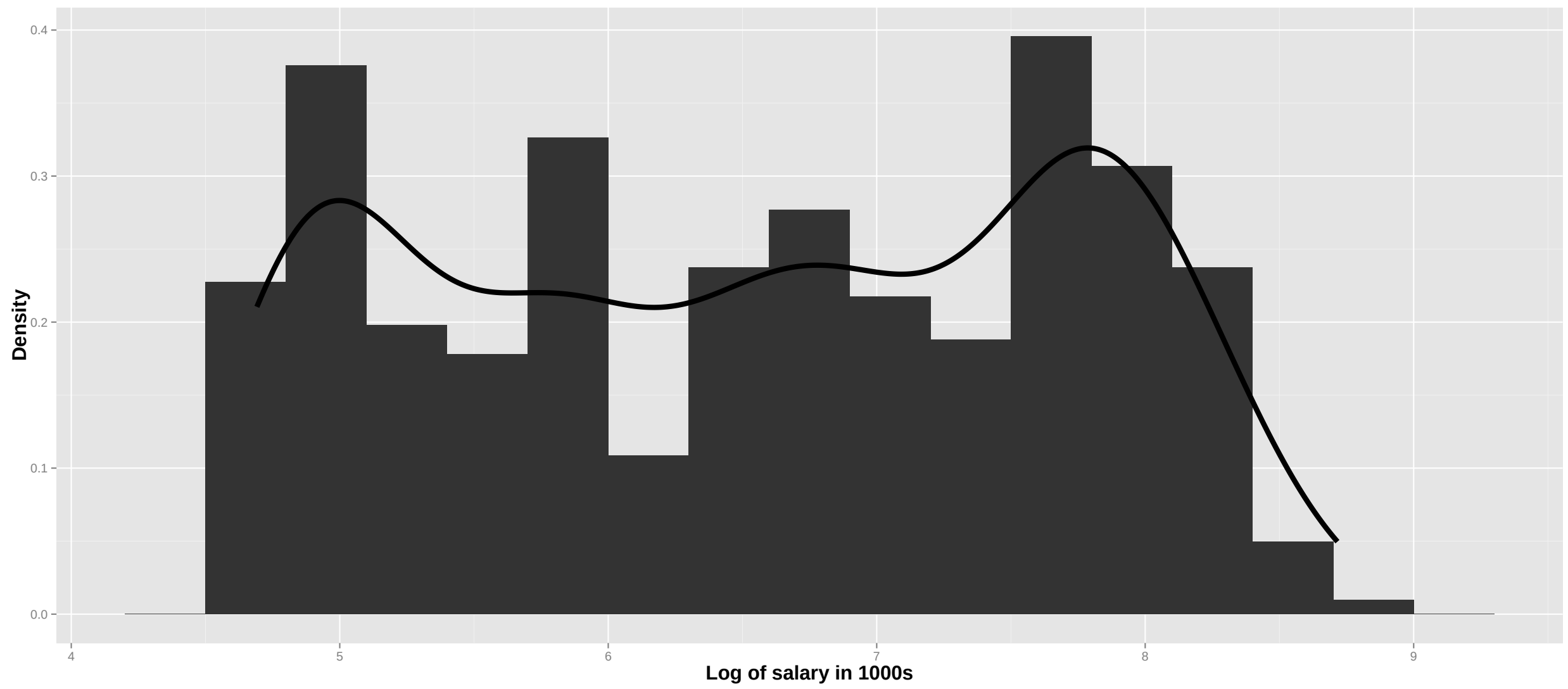

(b)

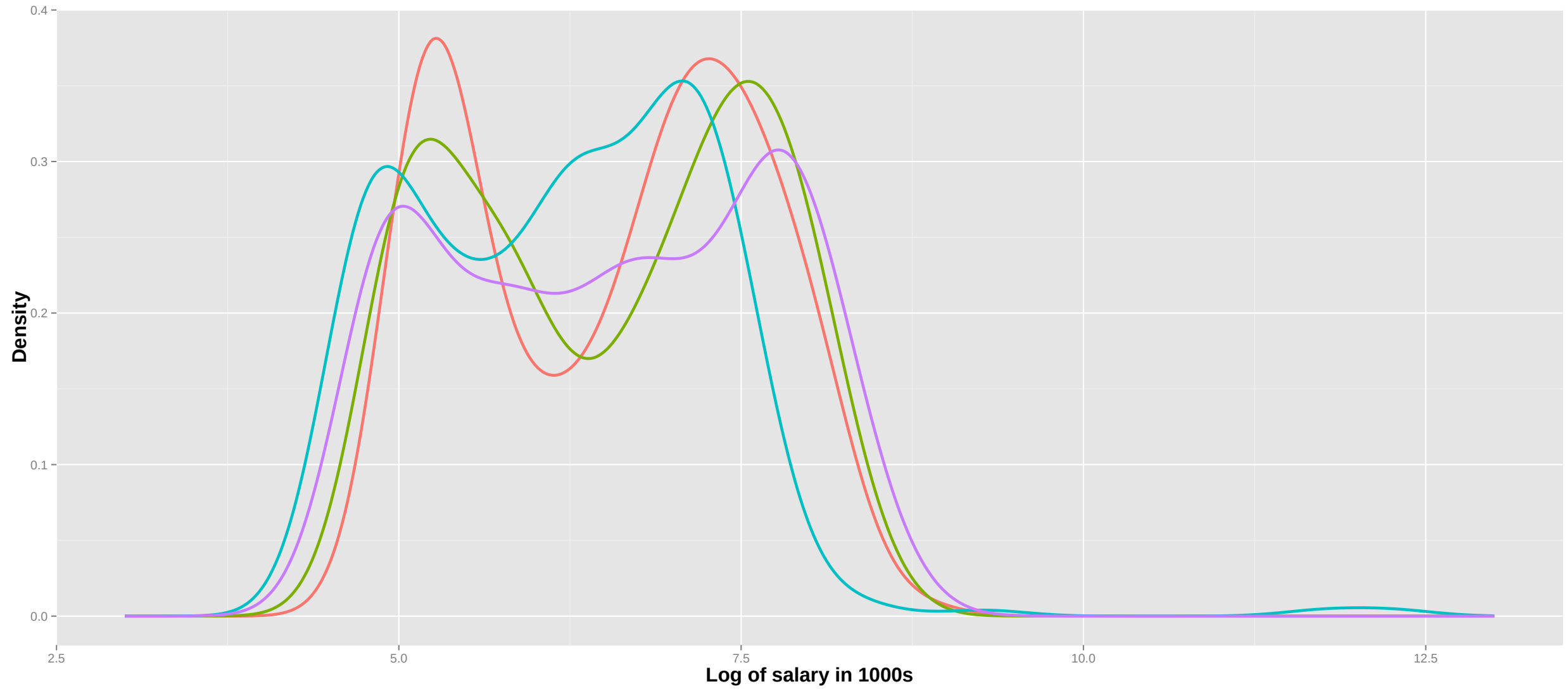

Figure 1: Summary of raw and predicted log(salary) for batters from Major League Baseball in years 1991/92. Panel (a) depicts a histogram of the log of salaries in 1000s for all 337 batters with the a fitted density line. Panel (b) shows the predicted densities of log salaries in 1000s from the stepwise-BIC (red), MIXLASSO (aqua), and L-MLR (green) from Table 3 with the observed data density (purple) also plotted for reference.

Table 4: Summary of stepwise-BIC (SW-BIC) and lasso-penalized mixture of linear regressions estimates from MLB data from the 2011-2015 seasons. Covariate acronyms are AB - At Bats, R - Runs, H - Hits, 2B - Doubles, 3B - Triples, HR - Home Runs, RBI - Runs Batted In, SB - Stolen Bases, CS - Caught Stealing, BB - Walks, SO - Strikeouts AVG - Batting Average, OBP - On Base Percentage, SLG - Slugging Percentage, OPS - OPS = OBP + SLG, WAR - Wins Above Replacement GP - Games Played, TPA - Total Plate Appearances, PIT - Number of Pitches, P/PA - Pitches Per Plate Appearance, TB - Total Bases, XBH - Extra Base Hits, HBP - Hit By Pitch, IBB - Intentional Walks, GDP - Grounded Into Double Plays, SH - Sacrifice Hits, SF - Sacrifice Flies, RC - Runs Created, RC27 - Runs Created Per 27 Outs, ISP - Isolated Power, SECA - Secondary Average, GB - Ground Balls, FB - Fly Balls, G/F - Ground Ball to Fly Ball Ratio, AB/HR - At Bats Per Home Run, BB/PA - Walks Per Plate Appearance, BB/K - Walk to Strikeout Ratio.

| Covariates | Season 2011 | | | Season 2012 | | | Season 2013 | | | Season 2014 | | | Season 2015 | | | Combined | | |
|---|---|---|---|---|---|---|---|---|---|---|---|---|---|---|---|---|---|---|
| | SW-BIC | Comp 1 | Comp 2 | SW-BIC | Comp 1 | Comp 2 | SW-BIC | Comp 1 | Comp 2 | SW-BIC | Comp 1 | Comp 2 | SW-BIC | Comp 1 | Comp 2 | SW-BIC | Comp 1 | Comp 2 |
| Intercept | 8.93 | 8.57 | 9.22 | 8.92 | 8.45 | 9.29 | 8.86 | 8.10 | 9.28 | 8.84 | 7.68 | 9.19 | 9.02 | 8.29 | 9.42 | 8.57 | 7.54 | 8.90 |
| AB | – | – | – | -10.84 | – | – | – | – | – | – | – | – | -12.44 | – | – | – | – | – |
| R | – | – | – | – | 0.10 | – | – | – | – | – | – | – | – | – | – | – | 0.15 | 0.10 |
| H | – | – | – | 28.54 | – | – | 0.50 | – | – | -3.10 | – | – | 25.11 | – | – | 0.28 | – | – |
| 2B | – | – | – | 3.70 | – | – | – | – | – | – | – | – | 2.93 | – | – | – | – | -0.15 |
| 3B | – | – | 0.12 | 2.65 | – | -0.06 | – | – | – | – | – | -0.10 | 1.63 | – | -0.19 | 0.21 | 0.15 | – |
| HR | – | – | – | 14.16 | – | – | 0.57 | – | – | – | – | – | 15.03 | – | – | 0.80 | – | – |
| RBI | – | 0.16 | 0.25 | – | – | – | – | – | 0.10 | – | – | 0.09 | – | 0.20 | – | 0.41 | – | 0.25 |
| SB | – | – | – | 2.88 | – | – | – | 0.07 | -0.01 | – | – | – | 1.78 | -0.01 | – | – | – | 0.11 |
| CS | -0.13 | – | -0.05 | -1.82 | – | -0.02 | – | – | – | – | – | – | -1.58 | – | -0.07 | – | -0.05 | – |
| BB | – | – | – | 9.40 | – | – | 0.36 | – | – | -2.45 | – | – | 12.15 | – | – | – | – | 0.06 |
| SO | – | – | – | -0.60 | – | – | -0.18 | – | – | – | – | – | – | – | – | -0.17 | – | -0.14 |
| AVG | – | – | – | – | – | -0.05 | – | -0.13 | 0.10 | – | – | – | – | – | – | – | – | – |
| OBP | – | – | – | – | – | 0.15 | – | – | 0.24 | 2.17 | – | – | -8.13 | – | – | – | – | – |
| SLG | – | – | – | -1.56 | – | – | – | – | – | – | – | – | – | – | – | – | – | – |
| OPS | – | – | – | – | – | – | -0.58 | – | – | – | – | – | 12.96 | – | – | – | – | – |
| WAR | -0.25 | – | -0.14 | – | – | – | – | – | – | – | -0.05 | – | – | -0.26 | – | -0.33 | -0.08 | -0.24 |
| GP | -0.18 | – | – | – | – | – | -0.48 | – | – | -0.40 | -0.04 | – | – | – | – | -0.31 | -0.03 | -0.35 |
| TPA | – | – | – | – | – | – | – | – | – | 2.90 | – | – | – | – | – | – | – | 0.49 |
| PIT | – | – | – | – | – | – | – | 0.02 | – | – | -0.16 | – | 3.49 | – | – | – | – | – |
| P/PA | – | – | – | – | – | – | – | – | – | – | – | – | -1.65 | – | – | – | – | – |
| TB | – | 0.06 | – | – | – | – | – | – | – | – | – | – | – | – | – | – | – | – |
| XBH | – | – | – | – | – | – | – | – | – | – | – | – | – | – | – | – | – | – |
| HBP | – | – | – | 2.31 | 0.09 | – | – | 0.08 | – | -0.41 | – | – | 2.65 | – | – | – | -0.08 | – |
| IBB | – | 0.06 | – | 0.15 | – | – | – | – | – | 0.26 | – | – | – | – | – | – | – | 0.12 |
| GDP | – | 0.19 | – | -3.23 | 0.21 | 0.17 | – | 0.31 | – | – | -0.17 | – | -3.90 | 0.08 | 0.08 | – | 0.08 | – |
| SH | -0.23 | – | -0.42 | – | – | – | – | – | -0.10 | – | – | -0.14 | – | -0.05 | – | – | -0.07 | -0.11 |
| SF | – | 0.06 | – | 0.29 | – | 0.15 | – | – | – | – | – | 0.08 | – | – | – | – | -0.05 | – |
| RC | – | 0.11 | – | -30.01 | – | – | – | – | – | – | – | – | -35.84 | – | – | – | – | – |
| RC27 | – | – | – | – | – | – | – | – | -0.21 | – | – | – | -2.90 | – | – | – | – | – |
| ISOP | – | – | – | – | – | – | – | – | – | – | – | – | -10.76 | – | – | -1.18 | – | – |
| SECA | 0.46 | – | – | 1.99 | – | – | – | 0.24 | – | – | – | – | 7.02 | – | – | 0.55 | 0.03 | – |
| GB | 0.44 | – | – | – | – | – | – | 0.15 | – | – | – | – | 1.50 | – | – | – | – | 0.06 |
| FB | 0.20 | – | – | – | 0.10 | – | – | – | -0.02 | – | – | – | – | – | 0.07 | – | – | – |
| G/F | – | – | 0.15 | – | – | – | – | – | -0.15 | – | – | – | -1.07 | – | -0.04 | – | -0.08 | – |
| AB/HR | – | – | – | – | -0.03 | – | – | – | – | -0.24 | – | – | – | – | -0.06 | -0.16 | – | -0.12 |
| BB/PA | – | – | -0.03 | – | – | – | – | – | – | – | – | – | – | – | – | – | – | – |
| BB/K | – | 0.06 | – | – | – | – | – | 0.11 | – | – | – | – | – | 0.14 | – | – | – | – |

(a) (b) (c) (d) (e) (f)

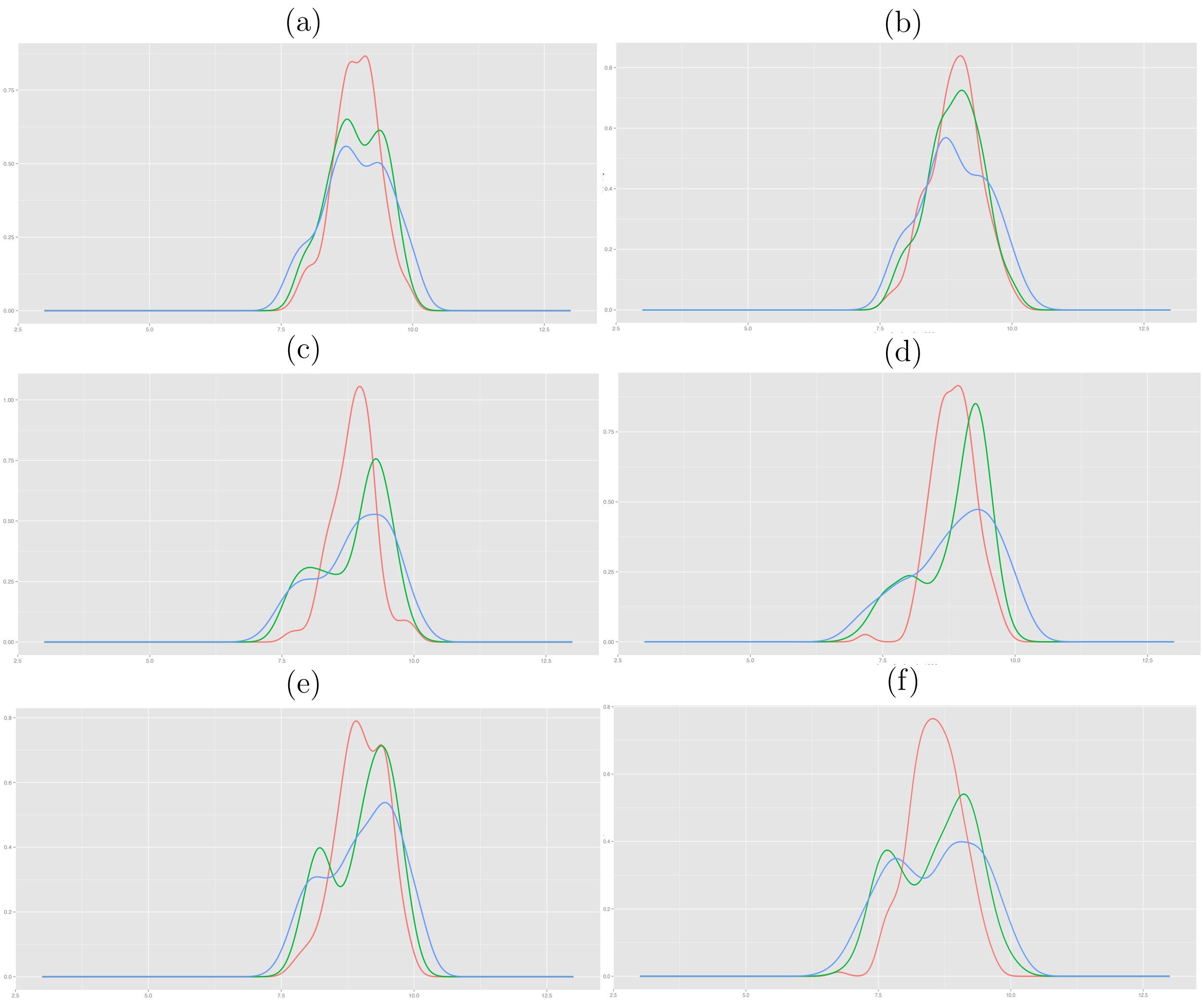

Log of salary in 1000s

Figure 2: Summary of densities for predicted and observed logged salaries (in 1000s of dollars) of batters from the MLB from seasons 2011-15. Panels (a) to (e) depict densities of logged salaries for the 2011-15 seasons with panel (f) depicting the predicted and observed densities of averaged logged salaries over all seasons. In each panel the blue line represents the density of observed logged salaries, the green the prediction from the L-MLR model, and red the predicted values from the stepwise-BIC linear model.

# 7 Discussion

In this work we have developed and implemented an algorithm that performs simultaneous model selection and estimation for a mixture of linear regressions model via lasso-penalized regression. The algorithm allows for coordinate-wise updates of parameters, and generates globally-convergent sequences of estimates that monotonically increase the penalized log-likelihood function, via the MM algorithm paradigm. We overcame the difficult step of correctly updating the mixing proportions for the mixture distribution by showing that the constrained system of equations can be converted to polynomial basis conversion problem that allows closed-form updates in the two, three and four component scenarios. For higher orders of components, fast polynomial root-finding algorithms can be exploited to find the mixing proportions in each iteration of the algorithm. We implemented the algorithm in a C++ program that is practically useful and is computationally efficient, allowing for large problems to be solved in reasonable time. We explored the use of a two-step procedure for optimising the $\boldsymbol{\lambda}$ vector by using the golden section search algorithm to find an initial optima on a predefined one-dimensional section, where $\lambda$ is fixed to be the same value for each element of $\boldsymbol{\lambda}$; the nonlinear optimization method of Nelder and Mead is then used to minimise the BIC criterion over the $\boldsymbol{\lambda}$ vector. This implementation is intuitively efficient and practical but does not lend itself to rigorous theoretical justification.

Several factors make penalized regression an ideal exploratory data analysis tool. One is the avoidance of the issues of multiple testing (Wu and Lange, 2008). Another is the generation of a predictive model that can be used to interpolate or extrapolate from the data. However, the generation of this predictive model requires (in our method) the minimization of the BIC over $\boldsymbol{\lambda}$, which we perform via numerical methods. The Nelder-Mead technique is a heuristic search method that can converge to non-stationary points and thus there is the potential that $\boldsymbol{\lambda}$ may not be a minimum of the BIC objective function. This is a practical trade-off between the computational demand of exhaustive grid search, which is prohibitive when $g$ is large, or when the grid is fine, and the potential location of a non-stationary point. We also assume that the location of a shared $\lambda$ via golden section search, is a good starting value for initializing the simplex of the Nelder-Mead method. Although the properties of Nelder-Mead are not known, to the best of our knowledge, we are the first to consider any sophisticated optimization over the $\boldsymbol{\lambda}$ vector in a mixture scenario. Although lasso-constrained $l_2$-regression is consistent, parameter estimates are biased toward zero in small samples (Wu and Lange, 2008). For this reason, once we have identified the active parameters for a given value of the tuning vector $\boldsymbol{\lambda}$, we re-estimate $\boldsymbol{\theta}$, ignoring the lasso penalty. As mentioned by Wu and Lange (2008), failure to perform this step leads to the inclusion of irrelevant predictors in the selected model.

In future work, the computational efficiency of the algorithm will be investigated with possible significant improvements to be expected with the implementation of a coordinate-descent algorithm (Wu and Lange, 2008), especially for scenarios where $p \gg n$. Like our MM algorithm, coordinate descent algorithms allow for an avoidance of matrix operations. Furthermore, they also permit the ability to test for whether costly operations for a predictor are required, which reduces updates to only those contributing predictors for a given $\boldsymbol{\lambda}$. This is in contrast to the current algorithm, which requires all predictors to be updated in each iteration. We reiterate that the presented algorithm is publicly available `https://github.com/lukelloydjones/` with source code and program compiled for Max OS X available. It is our intention, that with a more computationally efficient algorithm and implementation, to deploy our methodology on heterogeneous population problems in genomics. Our simulations and applications demonstrate that the modeling of heterogeneity in regression data is important for accurate prediction, as demonstrated by comparisons against stepwise-BIC. The solution to the mixing proportion polynomial problem is not only applicable in the L-MLR context, but also in other lasso-constrained mixture of regressions models, such as the logistic regressions model (also suggested in Khalili and Chen (2007)) or for Laplace mixtures (Nguyen and McLachlan, 2016) with lasso constraints.

# 8 Appendices

## A. Derivation of Theorem 2

Let

$$P_q(\xi;\ \boldsymbol{c}) = c_0 + c_1\xi + ... + c_q\xi^q \tag{22}$$

be a monomial basis form polynomial, with coefficients $\boldsymbol{c} = (c_0, ..., c_q)^T \in \mathbb{R}^{q+1}$ and $q \in \mathbb{N}$. Alternatively, we can write any polynomial in Newton basis form

$$P_q(\xi;\ \boldsymbol{d}, \boldsymbol{\delta}) = d_0 N_0(\xi;\ \boldsymbol{\delta}) + d_1 N_1(\xi;\ \boldsymbol{\delta}) + ... + d_q N_q(\xi;\ \boldsymbol{\delta}), \tag{23}$$

where $N_0(\xi;\ \boldsymbol{\delta}) = 1$ and $N_i(\xi;\ \boldsymbol{\delta}) = \prod_{j=0}^{i-1}(\xi - \delta_j)$, where $\boldsymbol{d} = (d_0, ..., d_q)^T \in \mathbb{R}^{q+1}$ and $\boldsymbol{\delta} = (\delta_0, ..., \delta_q)^T \in \mathbb{R}^{q+1}$.

Consider that we can write the left-hand side of (14) as

$$Q_g(\xi) = Q_g^*(\xi) - \sum_{i=1}^{g} Q_{g-1}^{(i)}(\xi),$$

where $Q_g^*(\xi) = \prod_{i=1}^{g}(\xi - b_i)$ and $Q_{g-1}^{(i)}(\xi) = a_i \prod_{j\neq i}(\xi - b_j)$, for each $i = 1, ..., g$. We firstly seek to write $Q_g^*(\xi)$ in form (22). We notice that $Q_g^*(\xi)$ can be written as $P_g(\xi;\ \boldsymbol{d}^*, \boldsymbol{\delta}^*)$ in form (23) by setting $d_g^* = 1$, $d_i^* = 0$ for $i \neq g$, $\delta_i^* = b_{i+1}$ for $i = 0, ..., g-1$, and arbitrarily setting $\delta_g \in \mathbb{R}$.

Let

$$\boldsymbol{L}_q(\boldsymbol{\delta}) = \begin{bmatrix} 1 & & & & \\ H_1(\delta_0) & 1 & & & \\ H_2(\delta_0) & H_1(\delta_0, \delta_1) & 1 & & \\ \vdots & \vdots & \ddots & \ddots & \\ H_q(\delta_0) & H_{q-1}(\delta_0, \delta_1) & \cdots & H_1(\delta_0, ..., \delta_{q-1}) & 1 \end{bmatrix},$$

where $H_i(\delta_0) = \delta_0^i$, $H_1(\delta_0, ..., \delta_i) = \sum_{j=0}^{i}\delta_j$, $H_i(\delta_0, \delta_1) = \sum_{j=0}^{i}\delta_0^j\delta_1^{i-j}$, and

$$H_i(\delta_0, ..., \delta_j) = (\delta_0 - \delta_j)^{-1}\left[H_{i+1}(\delta_0, ..., \delta_{j-1}) - H_{i+1}(\delta_1, ..., \delta_j)\right].$$

By the transformation theorem of Gander (2005), any polynomial of form (23) can be written in form (22) via the coefficient conversion formula

$$\boldsymbol{c} = \left[\boldsymbol{L}_q^T(\boldsymbol{\delta})\right]^{-1}\boldsymbol{d}. \tag{24}$$

Using (24), we can write $Q_g^*(\xi)$ as $P_g(\xi;\ \boldsymbol{c}^*)$ in form (22) by computing $\boldsymbol{c}^* = \left[\boldsymbol{L}_g^T(\boldsymbol{\delta}^*)\right]^{-1}\boldsymbol{d}^*$.

We now write each $Q_{g-1}^{(i)}(\xi)$ as $P_{g-1}\left(\xi;\ \boldsymbol{d}^{(i)}, \boldsymbol{\delta}^{(i)}\right)$ in form (23) by setting $d_{g-1}^{(i)} = a_i$, $d_j^{(i)} = 0$ for $j \neq g-1$, $\delta_j^{(i)} = b_{j+1}$ for $j = 0, ..., i-1$, $\delta_j^{(i)} = b_{j+2}$ for $j = i, ..., g-2$, and arbitrarily setting $\delta_{g-1} \in \mathbb{R}$. Using (24), we can write $Q_{g-1}^{(i)}(\xi)$ as $P_{g-1}\left(\xi;\ \boldsymbol{c}^{(i)}\right)$ by computing $\boldsymbol{c}^{(i)} = \left[\boldsymbol{L}_{g-1}^T\left(\boldsymbol{\delta}^{(i)}\right)\right]^{-1}\boldsymbol{d}^{(i)}$. We thus have

$$
\begin{aligned}
Q_g(\xi) &= P_g(\xi;\ \boldsymbol{c}^*) - \sum_{i=1}^{g} P_{g-1}\left(\xi;\ \boldsymbol{c}^{(i)}\right) \\
&= \sum_{j=0}^{g-1}\left(c_j^* - \sum_{i=1}^{g} c_j^{(i)}\right)\xi^j + c_g^*\xi^g. \qquad (25)
\end{aligned}
$$

# Supplementary figures

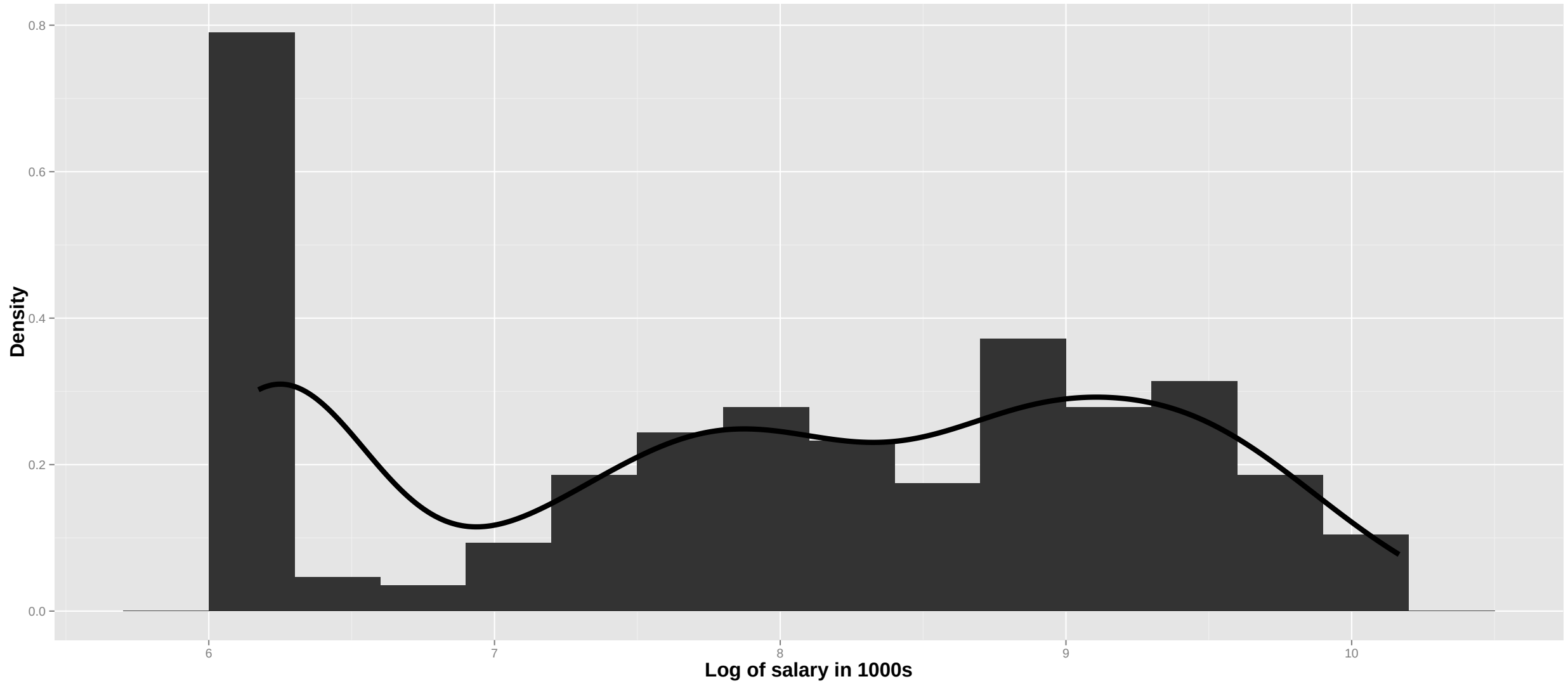

Figure S1: Histogram of averaged observed logged salaries (in 1000s of dollars) of batters from the MLB for seasons 2011-15. Figure shows salaries for 287 MLB batters with the solid line representing the density fit to the distribution.

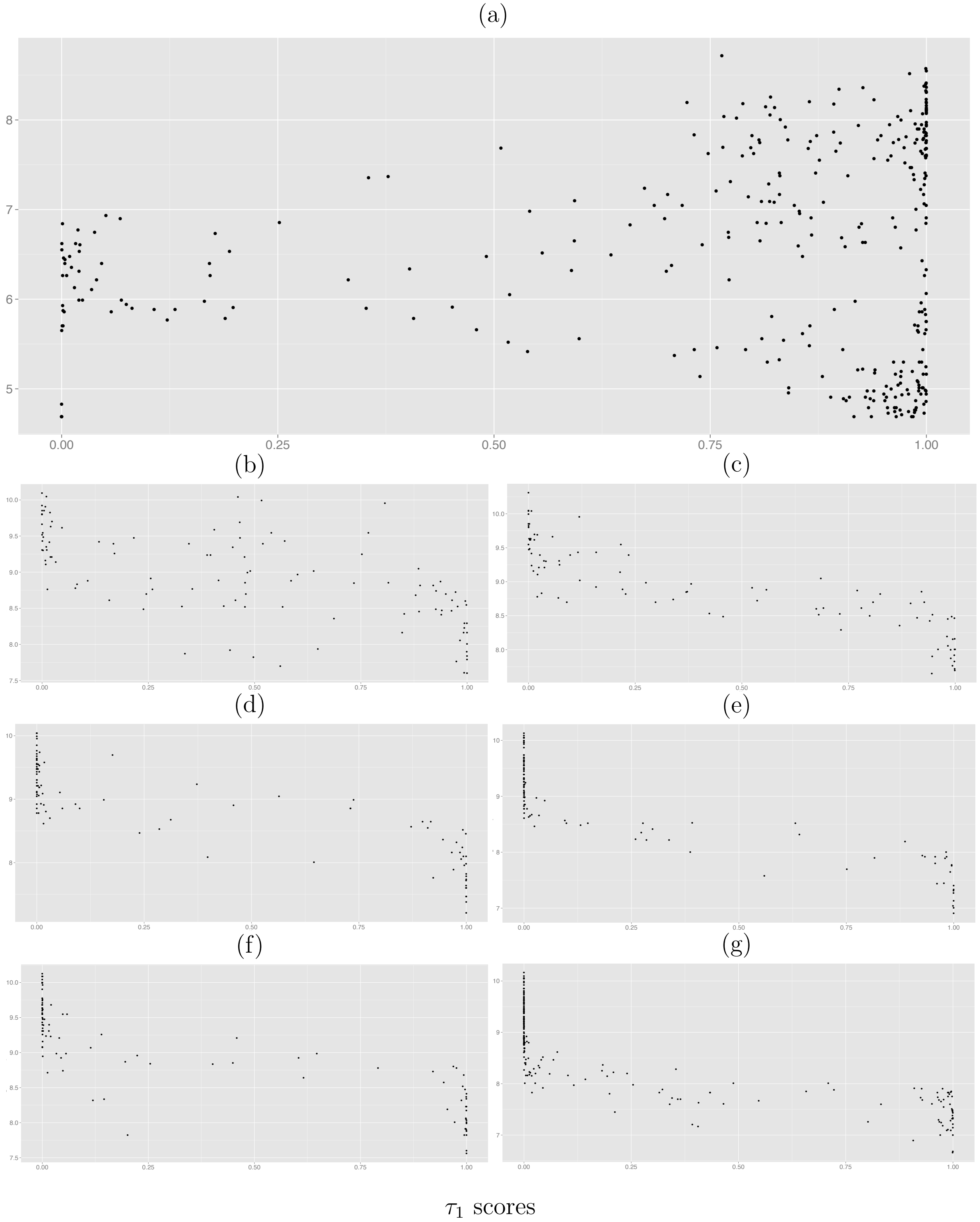

Figure S2: Summary of posterior probabilities versus log of salary in 1000s of dollars. Panels (a) through (g) depict results from MLB seasons 1991/92, 2011, 2012, 2013, 2014, 2015, and the combined analysis across seasons 2011 to 2015 respectively.

# Acknowledgements

The authors would like to thank the Australian Research Council for the research assistance and scholarships that funded this research.

# References

1000 Genomes Project Consortium and others (2012). An integrated map of genetic variation from 1,092 human genomes. *Nature 491*, 56–65.

Abrams, R. (2010). *The money pitch: Baseball free agency and salary arbitration.* Temple University Press, Philadelphia.

Brown, D. T., C. R. Link, and S. L. Rubin (2015). Moneyball after 10 years how have major league baseball salaries adjusted? *Journal of Sports Economics*, 1527002515609665.

Buhlmann, P. and S. van de Geer (2011). *Statistics for High-Dimensional Data.* New York: Springer.

De Veaux, R. D. (1989). Mixtures of linear regressions. *Computational Statistics and Data Analysis 8*, 227–245.

Dempster, A. P., N. M. Laird, and D. B. Rubin (1977). Maximum likelihood from incomplete data via the EM algorithm. *Journal of the Royal Statistical Society Series B 39*, 1–38.

DeSarbo, W. S. and W. L. Cron (1988). A maximum likelihood methodology for clusterwise linear regressions. *Journal of Classification 5*, 249–282.

Fullerton Jr, T. M. and J. T. Peach (2016). Major League Baseball 2015, what a difference a year makes. *Applied Economics Letters*, 1–5.

Gander, W. (2005). Change of basis in polynomial interpolation. *Numerical Linear Algebra with Applications 12*, 769–778.

George, E. I. (2000). The variable selection problem. *Journal of the American Statistical Association 95*, 1304–1308.

Greene, W. H. (2003). *Econometric Analysis*. Prentice Hall.

Grun, B. and F. Leisch (2007). Fitting finite mixtures of generalized linear regressions in R. *Computational Statistics and Data Analysis 51*, 5247–5252.

Hakes, J. K. and R. D. Sauer (2006). An economic evaluation of the Moneyball hypothesis. *The Journal of Economic Perspectives 20*, 173–185.

Hastie, T., R. Tibshirani, and J. Friedman (2009). *The Elements Of Statistical Learning*. New York: Springer.

Hui, F. K., D. I. Warton, S. D. Foster, et al. (2015). Multi-species distribution modeling using penalized mixture of regressions. *The Annals of Applied Statistics 9*, 866–882.

Hunter, D. R. and R. Li (2005). Variable selection using MM algorithms. *Annals of Statistics 33*.

Izenman, A. J. (2008). *Modern Multivariate Statistical Techniques*. New York: Springer.

Jones, P. N. and G. J. McLachlan (1992). Fitting finite mixture models in a regression context. *Australian Journal of Statistics 34*, 233–240.

Khalili, A. (2010). New estimation and feature selection methods in mixture-of-experts models. *Canadian Journal of Statistics 38*, 519–539.

Khalili, A. (2011). An overview of the new feature selection methods in finite mixture of regression models. *Journal of the Iranian Statistical Society 10*, 201–235.

Khalili, A. and J. Chen (2007). Variable selection in finite mixture of regression models. *Journal of the American Statistical Association 102*, 1025–1038.

Khalili, A. and S. Lin (2013). Regularization in Finite Mixture of Regression Models with Diverging Number of Parameters. *Biometrics 69*, 436–446.

Kiefer, J. (1953). Sequential minimax search for a maximum. *Proceedings of the American Mathematical Society 4*, 502–506.

Lange, K. (2013). *Optimization*. New York: Springer.

Lewis, M. (2004). *Moneyball: The art of winning an unfair game.* WW Norton & Company, New York.

Major League Baseball Players Association (2011). Agreement 2012-2016, between Major League Baseball and the Major League Baseball Players Association. *Attachment 46*, 265–276.

Marchi, M. and J. Albert (2013). *Analyzing baseball data with R.* CRC Press, Boca Raton, Florida.

McLachlan, G. J. and D. Peel (2000). *Finite Mixture Models.* New York: Wiley.

McNamee, J. M. (1993). A bibilography on roots of polynomials. *Journal of Computational and Applied Mathematics 47*, 391–394.

Nelder, J. A. and R. Mead (1965). A simplex algorithm for functional minimization. *Computer Journal 7*, 308–313.

Nguyen, H. D. and G. J. McLachlan (2015). Maximum likelihood estimation of Gaussian mixture models without matrix operations. *Advances in Data Analysis and Classification 9*, 371–394.

Nguyen, H. D. and G. J. McLachlan (2016). Laplace mixture of linear experts. *Computational Statistics & Data Analysis 93*, 177–191.

Nocedal, J. and S. J. Wright (2006). *Numerical Optimization.* New York: Springer.

Pan, V. Y. (1997). Solving a polynomial equation: some history and recent progress. *SIAM Review 39*, 187–220.

Quandt, R. E. (1972). A new approach to estimating switching regressions. *Journal of the American Statistical Association 67*, 306–310.

R Core Team (2015). *R: A Language and Environment for Statistical Computing.* Vienna, Austria: R Foundation for Statistical Computing.

Razaviyayn, M., M. Hong, and Z.-Q. Luo (2013). A unified convergence analysis of block successive minimization methods for nonsmooth optimization. *SIAM Journal of Optimization 23*, 1126–1153.

Schwarz, G. (1978). Estimating the dimensions of a model. *Annals of Statistics 6*, 461–464.

Scully, G. W. (1974). Pay and performance in major league baseball. *The American Economic Review 64*, 915–930.

Stadler, N., P. Buhlmann, and S. van de Geer (2010). $l_1$-penalization for mxture regression models. *Test 19*, 209–256.

Tibshirani, R. (1996). Regression shrinkage and selection via the Lasso. *Journal of the Royal Statistical Society Series B 58*, 267–288.

Venables, W. N. and B. D. Ripley (2002). *Modern Applied Statistics with S* (Fourth ed.). New York: Springer. ISBN 0-387-95457-0.

Wu, T. T. and K. Lange (2008). Coordinate descent algorithms for LASSO penalized regression. *Annals of Applied Statistics 2*, 224–244.

Yuan, M. and Y. Lin (2006). Model selection and estimation in regression with grouped variables. *Journal of the Royal Statistical Society Series B 68*, 49–67.

Zhou, H. and K. Lange (2010). MM algorithms for some discrete multivariate distributions. *Journal of Computational and Graphical Statistics 19*, 645–665.